\documentclass{article}
\usepackage[utf8]{inputenc}
\usepackage[legalpaper, margin=0.7in]{geometry}
\usepackage{amsmath,amssymb}
\usepackage{multicol}
\usepackage[usenames, dvipsnames]{xcolor}
\usepackage{graphicx,placeins}
\usepackage{hyperref}
\usepackage{siunitx}
\usepackage{cancel,units}

\usepackage{multibib}
\newcites{latex}{\LaTeX-Literature}

\title{Evidence for an intermediate-mass black hole from a gravitationally lensed gamma-ray burst}

\author{James Paynter\footnote{jpaynter@student.unimelb.edu.au, School of Physics, University of Melbourne, Parkville, Victoria, 3010, Australia}
	\and
	Rachel Webster\footnote{School of Physics, University of Melbourne, Parkville, Victoria, 3010, Australia}
	\and
	Eric Thrane
	\footnote{School of Physics and Astronomy, Monash University, Clayton, VIC 3800, Australia}\footnote{OzGrav: The ARC Centre of Excellence for Gravitational Wave Discovery, Australia}
}

\begin{document}
\maketitle

	{\bf
		If gamma-ray bursts are at cosmological distances, they must be gravitationally lensed occasionally~\citelatex{Paczynski_1986ApJ_cosmological_grbs,Paczynski__1987ApJ__GRB_microlensing}.
		The detection of lensed images with millisecond-to-second time delays provides evidence for intermediate-mass black holes, a population which has been difficult to observe.
		Several studies have searched for these delays in gamma-ray burst light curves, which would indicate an intervening gravitational lens~\citelatex{wambsganss_1993_apj,nowak_grossman_1994ApJ,Munoz:2016tmg,Ji_kovetz_grb_lensing_2018}.
		Among the $\sim10^4$ gamma-ray bursts observed, there have been a handful of claimed lensing detections~\citelatex{Hirose2006}, but none have been statistically robust.
		We present a Bayesian analysis identifying gravitational lensing in the light curve of GRB~950830.
		The inferred lens mass depends on the unknown lens redshift $z_l$, and is given by $(1+z_l)M_l = 5.5^{+1.7}_{-0.9}\times 10^4 $ M$_\odot$ (90\% credibility), which we interpret as evidence for an intermediate-mass black hole.
		The most probable configuration, with a lens redshift $z_l\sim 1$  and a gamma-ray burst redshift $z_s\sim 2$, yields a present day number density of $\approx 2.3^{+4.9}_{-1.6}\times10^{3} \text{ Mpc}^{-3}$ (90\% credibility) with a dimensionless energy density $\Omega_\textsc{IMBH} \approx 4.6^{+9.8}_{-3.3}\times10^{-4}$.
		The false alarm probability for this detection is $\sim0.6\%$ with trial factors.
		While it is possible that GRB~950830 was lensed by a globular cluster, it is unlikely since we infer a cosmic density inconsistent with predictions for globular clusters $\Omega_\textsc{GC} \approx 8 \times 10^{-6}$ at 99.8\% credibility.
		If a significant intermediate-mass black hole population exists, it could provide the seeds for the growth of supermassive black holes in the early Universe.
	}

	The evidence for a cosmological population of intermediate-mass black holes (IMBHs) is mounting.
	They have long been posited to reside in the cores of globular clusters.
	Dynamical friction in stellar clusters causes the most massive stars to sink to the bottom of the cluster's gravitational potential.
	Since 2004, simulations have indicated that, for small compact clusters, stellar mergers happens within the lifetime of giant stars~\citelatex{zwart_2004Nature}.
	Critically, these mergers occur before the stars go supernova and disturb the system, leading to a runaway collision and the formation of a $\sim10^3$ M$_\odot$ megastar.
	These short-lived monsters could seed IMBHs, which subsequently grow through accretion and mergers.
	Yet direct observational signatures of their existence are elusive.

	Their large mass puts the majority of IMBH mergers outside the sensitivity range of the current generation of gravitational-wave detectors.
	The Advanced Laser Interferometer Gravitational-wave Observatory (LIGO)~\citelatex{LIGO} and Virgo~\citelatex{Virgo} are sensitive to mergers with a total merger-product mass of $\lesssim\unit[400]{M_\odot}$.
	Furthermore, IMBHs are too small to be observed using the same techniques employed to detect supermassive black holes in galactic nuclei.
	They are either not massive enough, or live in a state of starvation, unable to accrete enough gas to power quasar-like emission.

	Astronomers are converging on the population of IMBHs from both ends of the black hole mass spectrum.
	Compact object mergers detected by LIGO-Virgo are uncovering a population of black holes edging closer  to the range traditionally reserved for IMBHs~\citelatex{Abbott_2019}, including the recent discovery of a 150 M$_\odot$ merger product~\citelatex{ligo_imbh}.
	From the other end, the lower limit on supermassive black holes in the nuclei of dwarf galaxies is descending~\citelatex{king_2020MNRAS}.
	There have been recent findings of compact objects with mass $\unit[10^{4}-10^5]{M_\odot}$ residing in galactic cores \citelatex{Tomoharu_2017NatAs,Takekawa_2020}.
	Observations of a tidal disruption event from the evisceration of a star by a black hole's tidal field suggest an IMBH resides in a star cluster on the outskirts of a barred lenticular galaxy~\citelatex{Lin_2018NatAs}.
	In this Letter, we provide evidence for an IMBH using gravitational lensing.
	Gravitational lensing is one of the few ways to directly constrain IMBH population statistics by providing an estimate for the number density of IMBHs.

	In \emph{strong} gravitational lensing, photon paths from a background source are distorted due to curved spacetime,
	producing multiple images.
	The relative fluxes and the difference between arrival times for each image can be used to infer the gravitational structure of the lens. In the case of a compact lens, the mass can be directly determined up to a redshift factor.
	The fraction of distant sources which experience multiple-imaging
	is directly proportional to the dimensionless energy density of compact lenses, $\Omega_\text{lens}\equiv\rho_\text{lens}/\rho_c$ \citelatex{Press_Gunn_1973_APJ} where $\rho_\text{lens}$ is the energy density of lenses while $\rho_c$ is the critical energy density required for a flat Universe.
	This fraction is independent of the lens mass, $m_\text{lens}$.
	Strong lensing is accompanied by an overall magnification, typically a factor of a few in flux.
	This allows us to probe more distant sources, or sources which would otherwise be too faint to detect.

	Gamma-ray bursts (GRBs) are extremely luminous bursts of $\gamma$-rays, with peak energies of $\unit[100-300]{keV}$.
	They are thought to be generated by the rapid infall of material onto a nascent stellar mass black hole, formed through either a collapsar supernova or a compact object merger.
	Some of the accreting material is launched in ultra-relativistic, bipolar jets along the rotation axis.
	A fraction of this outflow is converted into electromagnetic radiation, which is Lorentz-boosted into $\gamma$-rays.
	The cosmological nature of GRBs is well established by both the isotropy of observed events~\citelatex{Fishman_Meegan_1992_conference} and redshift measurements of their optical afterglow~\citelatex{costa_frontera_nature_1997}.
	A cosmological origin implies that at least some fraction of the GRB population must be strongly lensed~\citelatex{Paczynski_1986ApJ_cosmological_grbs}.

	Gamma-ray detectors, unlike those used for optical or infrared astronomy, have comparatively poor angular resolution, but good temporal resolution.
	Thus,  we do not expect to resolve a gravitationally lensed image pair in $\gamma$-rays.
	However, a time delay between the two images, resulting from the differences in geometric path and relative differences in gravitational field strength, can be observed.
	The photons which travel a longer distance arrive first, as the shorter path traverses deeper into the gravitational potential well of the lens where time dilation is stronger.
	The gravitationally retarded image is dimmer than the first image.
	The observational signature of such an event is thus an initial $\gamma$-ray pulse followed by a duplicate ``echo.''
	The duration of the time delay between the burst and the echo is predominantly determined by the mass of the gravitational lens, but also by the alignment of the $\gamma$-ray source with respect to the observer-lens line-of-sight.
	For a point-mass lens~\citelatex{narayan_wallington_1992_apj,Mao_1992_APJ_389,Krauss1991},

	\begin{equation}
		(1+z_\text{l})M = \frac{c^3\Delta t}{2G}\left(\frac{r-1}{\sqrt{r}} +\ln r\right)^{-1}.
		\label{eq:mass_redshift}
	\end{equation}
	Here, $\Delta t$ is the time delay, $r$ is the ratio of the fluxes, and $(1+z_\text{l})M$ is the redshifted lens mass.
	By measuring $\Delta t$ and $r$ we can infer the redshifted mass $(1+z_\text{l})M$.

	The total number of observed GRBs is of order $10^4$.
	We analyse the BATSE dataset as it is the largest available single dataset at $\sim2,700$ bursts.
	We include both long and short GRBs in our study.
	For a burst and echo to occur within the same BATSE light curve, we require a time delay of $\lesssim \unit[240]{s}$.
	The minimum detectable time delay is determined by the width of the $\gamma$-ray pulse; if the delay time is too short, the two images merge into one.
	For long GRBs, the minimum detectable time delay is $\approx \unit[1]{s}$, and for short bursts, it is $\approx$ $\unit[40]{ms}$.
	This range of time delays corresponds to a lens mass range of approximately $\unit[10^2- 10^7]{\text{ M}_\odot}$~\citelatex{Blaes1992,Mao_1992_APJ_389}.

	We identify preliminary lensing candidates with an auto-correlation analysis~\citelatex{geiger_schneider1996MNRAS,Hirose2006}.
	We utilise the four available broadband energy channels of BATSE burst data independently.
	The equivalence principle dictates that all wavelengths of light are equally affected by gravitational fields.
	This implies two constraints:
	the time delay is independent of the photon energy, and the gravitational magnification of each image is identical for every wavelength.
	Once we have identified candidates, we employ Bayesian model selection to determine the Bayesian odds comparing the lensing hypothesis to the no-lensing hypothesis.
	Our unified framework simultaneously provides the detection significance while estimating the lensing parameters, which we use to infer the lens mass.
	To model GRB pulses, we employ the fast-rise exponential-decay (FRED) model~\citelatex{1996ApJ_Norris_Nemiroff}.
	Details are provided in the supplementary material.

\begin{figure}
		\centering
		\includegraphics[width=1\textwidth]{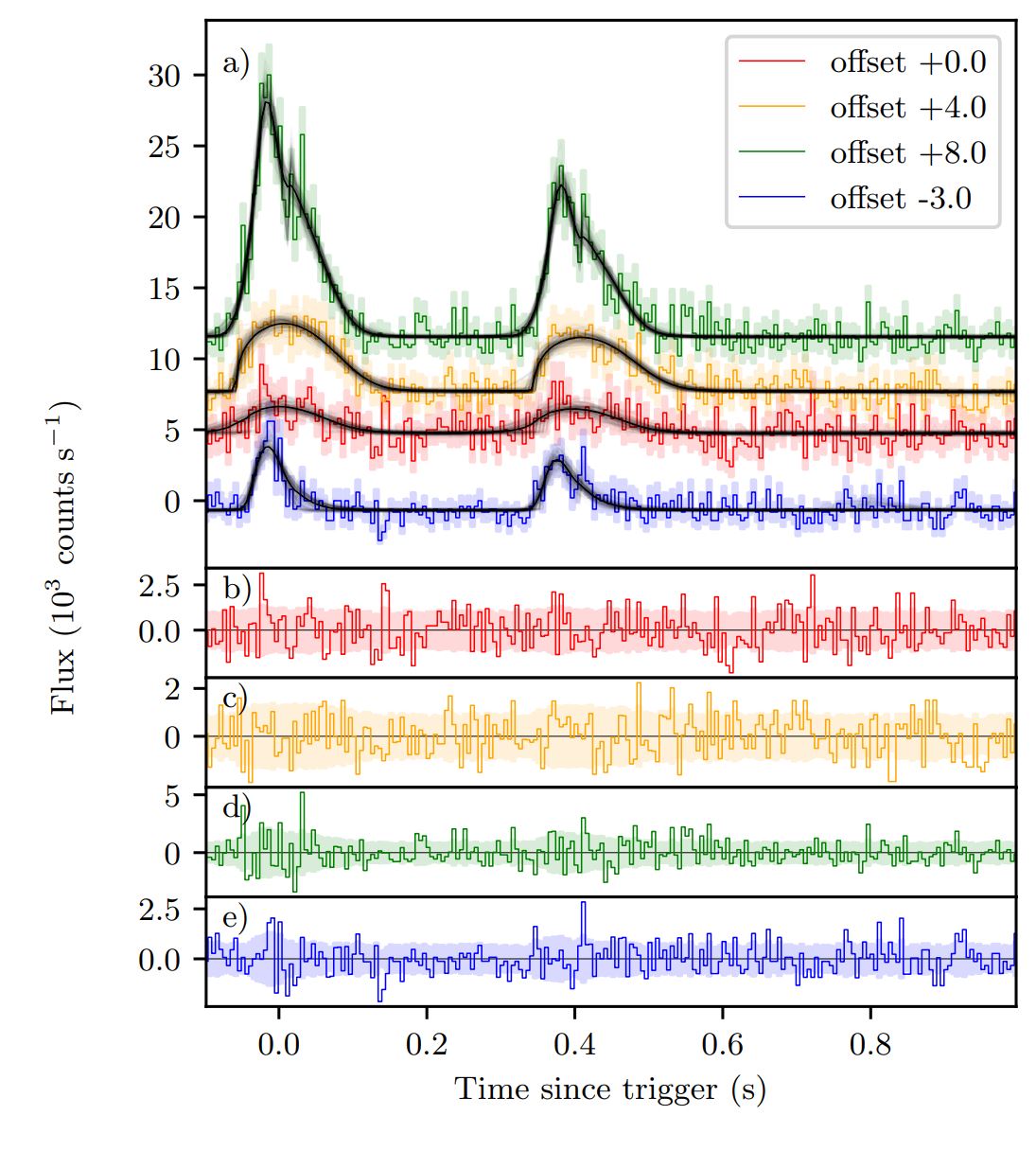}
		\caption{
			{\bf The gravitationally lensed $\gamma$-ray Burst, BATSE trigger 3770 -- GRB~950830.}
			a) The light curve is the pre-binned 5ms \texttt{tte BFITS} data.
			Each colour indicates a different energy channel, red: $20-60$ keV, yellow: $60-110$ keV, green: $110-320$ keV, blue: $320-2,000$ keV.
			The coloured shaded regions are the 1-$\sigma$ statistical errors of the $\gamma$-ray count data.
			The solid black curves are the posterior predictive curves, the mean of $\gtrsim 60,000$ fits.
			As they are the mean of $\gtrsim 60,000$ fits, the burst and echo image-fits need not be the same, even though the individual fits which make up the mean are identical.
			Over-plotted in fainter black are curves of 100 random parameter draws from the $\gtrsim 60,000$ total posterior sample sets.
			b-e) The lower panels are the difference between the true light curve and the posterior predictive curve.
			The shaded regions have been transformed in the same fashion.
			Note that 68\% of the peaks and troughs in the residual should lie within the corresponding shaded region for a good fit.
			The individual pulses that make up the combined fits are shown in Extended Data Figs. 1, 2, 3, and 4 for channels 1, 2, 3, and 4 respectively.
		}
		\label{fig:lightcurve}
	\end{figure}

	We uncover one statistically significant gravitational lensing candidate: GRB~950830 (BATSE trigger 3770)---a short $\gamma$-ray burst.
	The light curve for this burst is shown in Fig.~\ref{fig:lightcurve} with the reconstructed curve of the best-fit model plotted in black.
	The black curve is created by taking the mean of the curves drawn by each of the $\gtrsim 60,000$ posterior sample sets at each time bin.
	We find that each individual pulse is best fit by a variation of the FRED pulse model plus a sine-Gaussian function.
	We analyse the four available energy channels independently and find that the lensing hypothesis is preferred in each channel with $\ln$ Bayes Factors ($\ln(\text{BF})$) between $0.5 - 7$.
	Adding the $\ln(\text{BF})$s from each of the channels, we find the total $\ln(\text{BF})=12.9$ ($\log_{10}\text{BF}=5.6$) in favour of lensing, indicating strong statistical support for the lensing hypothesis.
	A $\ln(\text{BF})$ of eight is considered ``strong evidence'' in support of one model over the other~\citelatex{intro}. Detailed fits are shown in Extended Data Figs.1-6, including an example of a 'double' burst which is not a lens (Extended Data Figs. 7-8).

	Assuming a point-mass deflector, the marginalised posterior distributions for time delay and magnification ratio of this lensing event in Fig.2 can be used in conjunction with equation \eqref{eq:mass_redshift} to infer a redshifted lens mass of (Fig.3)
	\begin{equation}
		(1+z_\text{l})M_\text{l}\sim 5.5^{+1.7}_{-0.9} \times 10^4 \text{ M}_\odot.
	\end{equation}
	There are three astrophysical objects in this mass range, which might serve as a lens: globular clusters, dark matter halos, and black holes.
	A gravitational lens is well approximated as a point mass if most of its mass is contained within the region bound by the two lensed images where they bisect the cosmological plane of the lens.
	Taking instead an isothermal mass distribution as the gravitational lens, and integrating over all $z_\text{l}, z_\text{s}$, we find a lens velocity dispersion of $\sim\unit[4]{km\, s^{-1}}$.
	From simulations, this dispersion is associated with an Navarro–Frenk–White (NFW) profile of mass $\sim  10^5$ M$_\odot$~\citelatex{stu_chat}.
	Globular clusters follow a similar mass--velocity dispersion scaling~\citelatex{Baumgardt_2018MNRAS}.
	In either framework then, either a singular point mass, or a self-gravitating isothermal sphere, we have a consistent measurement for the mass.

	Dark matter halos are numerous, and their number density can be calculated using the Press-Schechter formalism.
	However, each has a negligible contribution to lensing cross-section, as NFW mass distributions typically have cores, which are not sufficiently massive to produce multiple images.
	Globular clusters are compact enough to produce multiple images, but there are not many of them.
	Assuming that the Milky Way's $\sim 200$ globular clusters are typical, and that the Milky Way formed from an overdensity of approximately $\unit[20]{Mpc^3}$, then the number density of globular clusters is approximately $\unit[10]{Mpc^{-3}}$, giving $\Omega_\textsc{gc}(10^5$ M$_\odot)\sim 8 \times 10^{-6}$---significantly lower than the mean density implied by GRB~950830.

	Following  \citelatex{Press_Gunn_1973_APJ}, we use the optical depth to estimate the cosmological density, $\Omega \sim \tau\left(\left<z_s\right>\right)$.
	Assuming that BATSE $\gamma$-ray bursts have a mean redshift of two, the IMBH energy density is
	\begin{equation}
		\Omega_\textsc{imbh}\left(M\sim10^{4-5} \text{ M}_\odot, \left<z_s\right>\sim2 \right)\approx 4.6^{+9.8}_{-3.3}\times10^{-4}.
	\end{equation}
	The present day number density of IMBHs is
	\begin{equation}
		n_\textsc{imbh}\left(M\sim10^{4-5} \text{ M}_\odot\right)\approx 2.3^{+4.9}_{-1.6}\times \unit[10^3]{Mpc^{-3}} ,
	\end{equation}
	(90\% credibility), where we have assumed a lens redshift of $z_l\sim1$.
	The uncertainty is from Poisson counting statistics.
	Thus, there should be approximately $\approx 4.6_{-3.2}^{+9.8}\times 10^4$ in the neighbourhood of the Milky Way.
	There are approximately $10^8$ stellar-mass black holes in the Milky Way~\citelatex{Elbert_2018MNRAS}.
	Assuming all stellar mass black holes are bound to galaxies, and with $n_\text{gal}\sim\unit[0.04]{Mpc^{-3}}$, then the number density of stellar mass black holes is $n_\text{stellar}\sim\unit[10^{7}]{Mpc}^{-3}$.
	Our result for the IMBH density is consistent with the stellar mass black hole density assuming that number density scales as $\sim M^{-1}$.
	Note that the mean redshift of Swift short GRBs is $\left<z_s\right>\approx0.8$. If the GRBs in the BATSE sample had the same mean, then the inferred cosmological density $\Omega_\textsc{imbh}$ would increase by about an order of magnitude.
	Extended Data Figs. 9-10 give results at different source and lens redshifts.

	Our estimate for $\Omega_\textsc{imbh}$ is consistent with the null result of other GRB lens searches~\citelatex{Ji_kovetz_grb_lensing_2018,hurley_2019ApJ}, which are sensitive to different lens masses.
	The Fermi and Konus-Wind catalogues are similar in size to the BATSE GRB catalogue, and
	there is $\sim50\%$ probability that these contain another GRB which is gravitationally lensed by an IMBH.
	In addition, due to the relatively flat GRB luminosity function~\citelatex{Blaes1992}, the uncertainty in $n_\textsc{imbh}$ derived from a single lensing event is more significant than the potential magnification bias.

	If this detection represents the first determination of the space density of IMBHs, then it may shed light on open questions in astrophysics.
	How are the supermassive black holes that power quasars so massive at high redshift?
	Are IMBHs gravitationally bound to galaxies?
	Do they have observational signatures in electromagnetic or gravitational radiation?
	What is their relationship to the globular cluster population?
	Are they the remnants of direct collapse of $10^{4-6}$ M$_\odot$ dark matter or baryonic clouds in the early Universe?
	The identification of additional lensing candidates in the GRB catalogues will confirm this result, and allow a more precise determination of $\Omega_\textsc{imbh}$.

	\section*{Acknowledgements}
	ET is supported through Australian Research Council Grant No. CE170100004 and No. FT150100281.
	The analysis software was run on The University of Melbourne's \texttt{Spartan} HPC system.
	This research has made use of data provided by the High Energy Astrophysics Science Archive Research Center (HEASARC), which is a service of the Astrophysics Science Division at NASA/GSFC and the High Energy Astrophysics Division of the Smithsonian Astrophysical Observatory.
	J.P. would like to acknowledge Stuart Wyithe, Michele Trenti, and Andrew Melatos for constructive comments in analysing and interpreting the data and results.
	J.P. would also like to thank Chris Shrader for assistance in understanding the BATSE instrumentation, and J. Michael Burgess for constructive feedback on \texttt{PyGRB} and the proper analysis of $\gamma$-ray data.

	\section*{Author Contributions}
	R.W. contributed to the initial planning of the project with later additions from J.P. and E.T.
	J.P. contributed the data analysis through the pulse-fitting software package \texttt{PyGRB} under the guidance of E.T.
	The manuscript was drafted by J.P. and E.T.
	J.P. and R.W. contributed the gravitational lensing calculations while E.T. contributed the Bayesian framework.
	J.P. and E.T. responded to questions and comments from the referees.
	All authors discussed the results and commented on the manuscript.

	\section*{Additional Information}
	Correspondence and requests for materials should be addressed to J.P.

	\section*{Competing Interests}
	The authors declare no competing financial interests.

	\section*{Methods}
	In this appendix we provide additional background on our methods.
	We start with a general overview of GRB lensing to place our research within the context of the wider field.
	From there, we describe our selection method for finding gravitationally lensed GRB candidates.
	We then discuss the statistics of photon counting in $\gamma$-ray astronomy.
	We construct a Bayesian framework with a model for the lensing signal and $\gamma$-ray background.
	We go on to discuss the validity and robustness of our results.
	We include calculations to determine the optical depth to lensing for a source population at mean redshift $\left<z_s\right>$, and provide evidence against the alternative hypothesis, that GRB~950830 was lensed by a globular cluster.
	We derive the uncertainty on our estimate for the number density of IMBH $n_\textsc{imbh}$.
	We also include an estimate of the false alarm probability, both with and without trial factors.
	Finally, we include a candidate identified by the autocorrelation detection algorithm but strongly rejected by our Bayesian analysis for illustrative purposes.

	\paragraph{A brief literature review.}
	Gravitational lensing studies of $\gamma$-ray bursts typically come in one of two flavours.
	There are autocorrelation studies, which search for echoes of the $\gamma$-ray burst within the same light curve.
	Then there are cross-correlation studies, which compare the light curve similarity of two separate GRB triggers on a per-bin basis.
	These are typically accompanied by positional coincidence statistics, which check if the GRBs have consistent source locations.
	Our study is of the first flavour.

	Traditional \emph{strong} gravitational lensing refers to the multiple imaging of, e.g., quasars due to galactic-mass gravitational lenses.
	Also known as macrolensing, fiducial image separations are on the order of arcseconds, with inter-image time delays of days to years depending on the lens geometry and source-lens alignment.
	Millilensing (or mesolensing) is loosely defined as gravitational lensing due to million solar mass objects~\cite{Marani1999ApJ}, and produces time delays of order seconds.
	The conversion between mass and time delay is
	\begin{equation}
		\Delta t \approx 50 \left(M_l / 10^6 M_\odot \right) \text{ seconds}
	\end{equation}
	for a Schwarzschild potential~\citelatex{Mao_1992_APJ_389}.
	In essence, millilensing fills the mass range between traditional strong lensing and microlensing whereby stars acting either alone or in unison in galaxies~\cite{Williams_wijers_1997,wyithe_2000MNRAS,lewis_2020} produce multiple images with $\sim$ microsecond time delays.
	At the more extreme end, nanolensing~\cite{walker_lewis_2003ApJ} ($\sim 10^{-6} - 10^{-1}$ M$_\odot$), picolensing ($\sim 10^{-12} - 10^{-7}$ M$_\odot$), and femtolensing~\cite{gould1992ApJ} ($\sim10^{-16} - 10^{-13}$ M$_\odot$) describe deflections and interference effects due to planetary and sub-planetary mass gravitational lenses.

	Autocorrelation probes millilensing echoes from $\sim10^2-10^6$~M$_\odot$ gravitational lenses.
	The minimum lens mass is determined by the temporal resolution of the instrument in addition to the variability timescale and duration of the burst.
	The upper mass limit is determined by the instrumental cut-off of data recording after the event trigger.
	Numerous autocorrelation searches of the BATSE database have been done using the summed $\unit[64]{ms}$ light curves~\cite{Nemiroff_Norris_1993,Ougolnikov2003},\citelatex{Hirose2006}.
	Autocorrelation has been used on the Fermi GBM and Swift BAT catalogues~\citelatex{Ji_kovetz_grb_lensing_2018}, with a null result for lenses masses of $10^1-10^3\text{ M}_\odot$.

	Cross correlation studies probe time delays equal to the difference in arrival time of the two bursts.
	The observation of the second image is inhibited by Earth occultation.
	This sets a minimum observable time delay, since $\gamma$-ray observatories typically have $\sim\unit[90]{min}$ orbital periods.
	Recent work on the Fermi GBM response shows how the observation conditions of a $\gamma$-ray burst significantly affect the inferred spectrum~\cite{Biltzinger2020}.
	The effects of detector angular response, energy response, atmospheric scattering, accumulated particle precipitation, cosmic background and Galactic sources such as the Sun or Crab Pulsar complicate cross-correlation studies.
	No lensing studies have looked for GRB lensing across multiple observatories due to the inherent difficulties comparing light curves from instruments with different energy responses.
	Cross-correlation studies of the BATSE database are also numerous~\cite{Nemiroff2000AIPConference,Li_Li_2014_ScienceChina}, in addition to Fermi GBM~\cite{davidson_bhat_conf}.
	A study of Konus-Wind GRBs searching for time delays of $\sim$ hours to $\sim25$ years (lens masses of $10^8-10^{13}\text{ M}_\odot$) was further augmented by the inclusion of spectral analysis~\citelatex{hurley_2019ApJ}.
	Another lens identification technique involves correlation of the cumulative light curve in three spectral dimensions~\cite{Bagoly2010AIPC}.
	Correlation-free approaches include a model-agnostic statistical method, which does take into account Poisson statistics~\citelatex{wambsganss_1993_apj}, and Fourier analysis methods~\citelatex{nowak_grossman_1994ApJ}.
	For a comprehensive review of the gravitational lensing of transient events, see~\cite{oguri_2019}\citelatex{hurley_2019ApJ}.

	\paragraph{Candidate selection.}
	The BATSE catalogue contains 2,704 triggered $\gamma$-ray bursts.
	Of these, 2,629 have \texttt{discsc bfits} ($\unit[64]{ms}$) light curves available for download.
	Furthermore, higher-time-resolution observations are available for 2,446 of these $\gamma$-ray bursts, with 2,435 existing as pre-binned \texttt{tte bfits} ($\unit[5]{ms}$) light curves.
	We carry out a preliminary auto-correlation search (described below) on both the \texttt{discsc bfits} and \texttt{tte bfits} pre-binned light curves.
	Candidate detections are followed up with further analysis.
	In total we carry out auto-correlation on 2,679 unique $\gamma$-ray bursts.

	Signal (auto)-correlation can be used to  measure the time delay of temporally overlapping signals of a gravitationally lensed system~\citelatex{geiger_schneider1996MNRAS}.
	We define the autocorrelation function (ACF) as
	\begin{equation}
		C(\delta t) = \frac{\sum_{j=0}^{n} I(t_j + \delta t)I(t_j)}
		{\sum_{k=0}^{N} I(t_k)^2},
		\label{eq:correlation}
	\end{equation}
	where the sum in the numerator is taken over the bins where the two signals overlap, and the sum in the denominator is taken over the entire input signal~\citelatex{geiger_schneider1996MNRAS, Hirose2006}.
	Here, $I(t_j)$ is the count rate at time bin $t_j$, $N$ is the total number of bins, and $n$ the total number of overlapping bin, where $j, k$ index these bins in the summations.

	We fit a $3^\text{rd}$ order Savitzky-Golay filter $F(\delta t)$ to the ACF.
	The dispersion $\sigma$ between the ACF and the fit $F(\delta t)$ is
	\begin{equation}
		\sigma^2 = \frac{1}{N}\sum_{j=0}^{N} [C(\delta t_j) - F(\delta t_j)]^2,
		\label{eq:dispersion}
	\end{equation}
	where $N$ is the total number of bins.
	We identify $3\sigma$ outliers as gravitational-lensing candidates~\citelatex{Hirose2006,Ji_kovetz_grb_lensing_2018}.
	Furthermore, we autocorrelate each of the four BATSE \texttt{LAD} energy channels, and perform the same filtering process.
	Gravitational lensing is achromatic for point sources, so we expect that each channel of a candidate lens GRB should autocorrelate with the same time delay.
	We check that candidates yield lensing signals in both the summed light curve and individual energy channels.

	\paragraph{Photon counting.}
	Photon counting is a Poisson process.
	High-energy satellites like BATSE accumulate photons at a series of discrete times, $t_1, t_2, ... t_n$.
	For BATSE, these photons are collected with a sampling frequency of $\unit[500]{kHz}$.
	In most cases, hardware limitations require that the BATSE photon arrival times (time-tagged events; TTE) are down-sampled before transmission to Earth.
	Only the shortest, moderately bright bursts are completely contained within the TTE photon-list data.
	Fortunately, this is the case for GRB~950830.
	For bursts not completely encoded in a TTE list, the counts are averaged into $\unit[64]{ms}$ bins before transmission, typically recorded for $\sim\unit[240]{s}$ after triggering.
	We ignore the fact that BATSE has a small dead time (of about one clock cycle) after each count as this is only becomes import for very bright bursts which saturate the detector.

	If we consider a single time stamp $t_i$, the likelihood of observing $N_i$ photons is given by Poisson counting statistics
	\begin{align}
		{\cal L}(N_i|\theta) = \frac{\lambda_{\theta,i}^{N_i} e^{-\lambda_{\theta,i}}}{N_i!} .
	\end{align}
	The expected number of photons, $\lambda_\theta$ is
	\begin{align}
		\lambda_{\theta,i} = \delta t_i \, R(t_i\theta) .
	\end{align}
	Here $\delta t_i$ is the sampling time and $R(t_i|\theta)$ is the photon rate (in units of photons per unit time) evaluated at time $t_i$ and given model parameters $\theta$.
	The sampling time, $\delta t_i$, is subscripted with index $i$ to account for the cases where the time resolution in the available data changes during an event.
	The rate can be written as a sum of signal $S$ (from the GRB) and background $B$
	\begin{align}
		R(t_i|\theta) = B + S(t_i|\theta) .
	\end{align}
	To first order, the background is constant, but the signal varies with time according to model parameters $\theta$.

	The likelihood of observing $\vec{N} = N_1, N_2, ...$ photons at times $t_1, t_2, ....$ is given simply by taking the product of the likelihood functions evaluated at different times
	\begin{align}
		{\cal L}(\vec{N}|\theta) = & \prod_i
		{\cal L}(N_i|\theta) .
	\end{align}
	It is easier, though, to work with the $\ln$ likelihood, which is
	\begin{align}
		\ln {\cal L}(\vec{N}|\theta) = & \sum_i
		\ln{\cal L}(N_i|\theta) \\
		= & \sum_i N_i\ln\Big(\delta t_i B +  \delta t_i S(t_i|\theta)\Big) \nonumber\\
		& -  \Big(\delta t_i B +  \delta t_i S(t_i|\theta)\Big) -\log(N_i!) .
	\end{align}

	\paragraph{Bayesian Inference.}
	There are two goals of Bayesian inference.
	The first is to derive posterior distributions $p(\theta|d)$ for our model parameters, which enables us to determine their credible intervals.
	The second goal is to calculate the Bayesian evidence $\mathcal{Z}$ for a set of models in order to do model selection.
	Bayes theorem,
	\begin{equation}
		p(\theta|d) = \frac{\mathcal{L}(d|\theta) \pi(\theta)}{\mathcal{Z}},
		\label{eq:bayestheorem}
	\end{equation}
	relates the posterior probability density $p(\theta|d)$ of model parameters $\theta$ given the observed data $d$, to a likelihood function $\mathcal{L}(d|\theta)$ and prior probability density $\pi(\theta)$.
	The likelihood function is a mathematical description of the probability of observing the data with the given model parameters.
	The priors are probability distributions for what we expect these parameters to be, which, in our case, are informed by the BATSE GRB population.
	The evidence, also called the marginal likelihood, is a normalisation factor which gives information about the quality of the fit of the model to the data averaged over parameter space, viz.
	\begin{equation}
		\mathcal{Z} = \int \mathcal{L}(d|\theta) \pi(\theta) d\theta .
		\label{eq:evidence}
	\end{equation}

	We define different models for
	\begin{align}
		S(t_i|\theta,M) ,
	\end{align}
	which we use to do Bayesian inference.
	The null model $M=M_\text{null}$ states that there is no lensing.
	The lens model $M=M_\text{lens}$ states that there {\em is} lensing.
	We adopt the Fast-Rise Exponential-Decay (FRED) pulse model, which is ubiquitous in GRB pulse-modelling:
	\begin{equation}
		S(t|A,\tau,\xi,\Delta) = A \exp \left[ - \xi \left(  \frac{t - \Delta}{\tau} + \frac{\tau}{t-\Delta}  \right)   \right] . \label{eq:FRED}
	\end{equation}
	Here, $A$ is a vertical y-scale factor, $\tau$ is a duration scaling parameter, and $\xi$ is an asymmetry parameter, which can be used to adjust the skewness of the pulse.
	A more generalised form of this model has additional exponents, $\gamma, \nu$, allowing for flatter/sharper peaks, viz.
	\begin{equation}
		S(t|A,\tau,\xi,\Delta,\gamma,\nu) = A \exp \left[ -\xi^\gamma \left(\frac{t - \Delta}{\tau}\right)^\gamma - \xi^\nu \left(\frac{\tau}{t-\Delta}\right)^\nu\right] . \label{eq:FREDx}
	\end{equation}
	We call this the extended FRED model, or FRED-X for short.
	An analytic normalisation exists for both the FRED and FRED-X models, which decouples the maximum height of the pulses from every parameter except $A$, such that $A$ is the maximum amplitude of the pulse.
	Structured pulses can be modelled as either multiple overlapping pulses, or by accounting for the residual structure with another parameterisation.
	Thus, a single channel FRED light curve requires $4n+1$ parameters, where the $+1$ corresponds to the constant background parameter $B$.
	For bursts with many pulses, our model is:
	\begin{align}
		S_\text{tot} = \sum_j S(t|\Delta_j, A_j, \tau_j, \xi_j) .
		\label{eq:multipulsemodel}
	\end{align}
	Our prior enforces
	\begin{align}
		\Delta_{j+1} > \Delta_j .
	\end{align}
	to ensure that we are not fitting the same pulse configuration in different permutations.

	The FRED model has proved a popular phenomenological fit due to its simplicity and \emph{a~posteriori} applicability to certain GRB progenitor models.
	Some authors have noted (and we confirm here) that there is a systematic residual structure, visible after subtracting the best-fit FRED and FRED-X models from a GRB light curve, indicating an imperfect fit~\cite{hakkila_2018}.
	We model these residuals using a sine-Gaussian wave packet
	\begin{equation}
		\text{res}(t)= A_\text{res} \exp \left[ - \left(\frac{t-\Delta_\text{res}} {\tau_\text{res}}\right)^2 \right] \cos\left(\omega t + \varphi \right),
		\label{eq:sinegaussian}
	\end{equation}
	as it is ubiquitous in physics and provides an adequate fit to the residual structure.
	The residual is part of our signal model, so we fit it simultaneously to any FRED pulses:
	\begin{equation}
		S(t|\theta) = S(t|\theta_\textsc{fred}) + \text{res}(t|\theta_\text{res}).
	\end{equation}

	The lens model is similar to the null model, but with two extra parameters used to describe the delayed signal:
	\begin{equation}
		S_\text{lens}(\theta_\text{lens}) = S(t|\theta_\text{null}) + r^{-1} \cdot S(t+\Delta t| \theta_\text{null})
		\label{eq:lens_model}
	\end{equation}
	where the lens parameter vector $\theta_\text{lens}$ subsumes the null parameter vector $\theta_\text{null}$ in addition to the time delay, $\Delta t$, and magnification ratio, $r$.
	The parameter $r$ reduces the amplitude of the delayed signal while the parameter $\Delta t$ describes the size of the delay.
	Thus, the lens model requires $4n+3$ parameters, where $n_\text{lens}$ is typically about half $n_\text{null}$.

	In order to determine which model is favoured, we calculate the Bayesian evidence for each model:
	\begin{align}
		{\cal Z}_\text{null} = & \int d\theta_\text{null} \, {\cal L}_\text{null}(\vec{N},\theta_\text{null})\pi(\theta_\text{null}), \\
		{\cal Z}_\text{lens} = & \int d\theta_\text{lens} \, {\cal L}_\text{lens}(\vec{N},\theta_\text{lens})\pi(\theta_\text{lens}) .
	\end{align}
	Here $\pi$ denotes a prior distribution and $\Vec{N}$ is the vector of photon counts.
	Once we have each evidence, we obtain the $\ln$ Bayes factor
	\begin{align}
		\ln(\text{BF}) = \ln{\cal Z}_\text{null} - \ln{\cal Z}_\text{lens} .
		\label{eq:bayes}
	\end{align}
	The Bayes factor is a statistically rigorous measure of which model the data prefer.
	A $\ln$ Bayes factor of eight, $\ln(\text{BF}) \gtrsim 8 $, is considered ``strong evidence'' in support of one model over the other~\citelatex{intro}.

	To perform parameter estimation and evidence calculations we use the \texttt{Bilby} Bayesian inference library~\cite{bilby_2019}.
	We employ nested sampling~\cite{skilling_2004,skilling_2006}, taking advantage of the multi-ellipsoid bounding method~\cite{feroz_2009_multinest}, with dynamically updated sampling points~\cite{higson_2019_dynamic_nested_sam}.
	Results tend to be unimodal, but we use multi-ellipsoid bounding regardless due to its flexibility and speed in the case of multimodal results.
	Some parameters recover bi-modal distributions, particularly in pulse start times $\Delta_j$, due to the pre-binning of the \texttt{BFITS} datatype analysed.

	\paragraph{Priors.}
	The priors are tabulated in Table~\ref{tab:priors}.
	We use log uniform priors on parameters, which typically vary by orders of magnitude and uniform priors for other parameters.
	The prior ranges are informed by the BATSE GRB population.
	We fit models to a large selection of isolated, individual pulses to develop intuition for the practical bounds of the priors.
	A more sophisticated analysis would employ hierarchical inference to infer the shape of the prior distributions, but we leave this for future work.
	The priors for quantities related to time ($\Delta, \Delta_\text{res}, \Delta t$) are taken to be uniform.
	The prior range depends on the particular trigger being investigated, since the GRB population is (bimodal) log-normal in duration over many orders of magnitude.
	Thus the priors must be chosen appropriately for a region of time that will contain all the pulses.
	Care is taken with the time delay prior to ensure that the extra pulses due to lensing are not occurring outside the light curve under investigation.
	Additionally, the prior forces the second pulse to occur after the first pulse.

	\begin{table*}[htb]
		\begin{center}
			\begin{tabular}{|l| r |r |l |l|}
				\hline
				parameter & minimum & maximum & type & units\\\hline
				$\Delta_i$ & $\dagger$ & $\dagger$ & uniform & seconds \\
				$\Delta_{i+1}$ & $\Delta_i$ & $\dagger$ & uniform & seconds\\
				$B$ & $10^{-1}$ & $10^{3}$ & log-uniform & counts / bin\\
				$A$ & $10^{0\hphantom{-}}$ & $10^{5}$ & log-uniform & counts / bin\\
				$\tau$ & $10^{-3}$ & $10^{3}$ & log-uniform & seconds\\
				$\xi $ & $10^{-3}$ & $10^{3}$ & log-uniform & -- \\
				$\gamma$ & $10^{-3}$ & $10^{3}$ & log-uniform & -- \\
				$\nu$ & $10^{-3}$ & $10^{3}$ & log-uniform & -- \\
				$\Delta_\text{res}$ & $\dagger$ & $\dagger$ & uniform & seconds \\
				$A_\text{res}$ & $10^{0\hphantom{-}}$ & $10^{3}$ & log-uniform & counts / bin\\
				$\tau_\text{res}$ & $10^{-3}$ & $10^{3}$ & log-uniform & counts / bin\\
				$\omega$ & $10^{-3}$ & $10^{3}$ & log-uniform & -- \\
				$\varphi$ & $-\pi$ & $\pi$ & uniform & radians \\\hline
			\end{tabular}
		\end{center}
		\caption{{\bf Priors used in this analysis.} Cells marked with $\dagger$ indicate that the prior bounds are chosen differently for each burst. Typically they are the start and end bins of the light curve. The prior for $\Delta_\text{res}$ has the same prior bounds as $\Delta$. See the methods paragraph Priors for more information.}
		\label{tab:priors}
	\end{table*}

	\paragraph{Results \& Interpretations.}
	For GRB~950830, we find that lensing is strongly preferred over two-pulse models.
	The two pulses in Fig.~\ref{fig:lightcurve} are so alike, that the data prefer a fit with a single set of pulse-parameters.
	Thus, a single pulse seen twice with delay time $\Delta t$, and reduced in brightness by some scaling factor $r$ is the preferred model compared to a more complex model with two completely independent pulses.
	We interpret this result as evidence that GRB~950830 was strongly lensed.

	A lensed FRED-X model with a sine-Gaussian residual is the preferred model; (see Eqs.~\eqref{eq:FREDx}, \eqref{eq:sinegaussian} for the signal model).
	There is only a single pulse, which is later repeated, so $j\in\{1\}$.
	The fits to the light curve for each channel are shown in Extended Data Figs.~(1-4).
	The median fits of the FRED-X pulses and sine-Gaussian pulses are plotted individually, and the sum of both is shown in the third panel of each figure.
	There are also 100 individual posterior draws for each model to show the breadth and multi-modality of the fits.
	Compared to the corresponding two-pulse model, the lens model is favoured by $\ln(\text{BF}) = 12.9$.
	We find that other lens models are similarly favoured.
	No matter which pulse model we implement, the lensing hypothesis is always favoured.
	The simplest model comparison comparing a single lensed FRED pulse to a null model with two independent FRED pulses, has the largest Bayes factor with $\ln(\text{BF}) = 24.5$.

	A model with more parameters is naturally penalised in Bayesian model selection by virtue of the larger region of prior volume explored when calculating the Bayesian evidence~\citelatex{intro}.
	We therefore ask: is the lens model favoured because an additional pulse simply adds such a great volume to our prior space?
	We investigate the effect of prior volume on the model selection to assure ourselves that we have not arrived at a spurious result.
	The extra parameters in the FRED-X model are $\gamma$ and $\nu$.
	We test priors on $\gamma$ and $\nu$ in the ranges, $(10^{-1},10^1)$, $(10^{-2},10^2)$, $(10^{-3},10^3)$ in both uniform and log-uniform spaces.
	We also include a narrow Gaussian prior centred on the values found in previous analysis (typically between 1/4 -- 4).
	In all, we study seven different prior volumes for two models in each of the four channels.
	The effect on the resultant model selection is minimal.
	Looking at the corresponding pairs of null and lens models which have the same priors, we find the $\ln(\text{BF})$ changes by $\sim1-2$.

	Channel 3 (green) exhibits the greatest variation in its Bayes factor, likely due to having the highest signal-to-noise ratio.
	For channel 4 (blue), the inclusion of a sine-Gaussian residual makes the two-pulse (null) model marginally preferred ($\ln(\text{BF}) \sim 1$) over a lens model.
	The $\ln(\text{BF})$s in favour of lensing are $\sim2-5$ per channel, depending on the prior.
	The total Bayes factor quoted in the main body of the paper assumes log-uniform priors for $\gamma$ and $\nu$ on $(10^{-1},10^1)$.
	In any case, the two parameters used to infer the lens mass are largely independent of the model or choice of priors.
	We find that our result is independent of the nested sampling package, eg. \texttt{Dynesty}~\cite{2020MNRAS_dynesty}, \texttt{Nestle}: \url{http://kylebarbary.com/nestle/}, used to run the analysis.

	As a sanity check, we apply our analysis not only to the pre-binned \texttt{tte\_bfits} data, but we also take the photon arrival time data (\texttt{tte\_list}) and run the analysis again on the counts incident at each triggered detector.
	There are three: detectors 5, 6, and 7.
	We bin the count data to 0.005 ms bins to match the pre-binned light curve and repeat our analysis.
	We find that there is no change to the model section.
	While individual Bayesian evidence factors and therefore model comparison Bayes factors fluctuate when considering different data, the lensing model is consistently preferred in each channel for each triggered detector.

	We also analyse the hardness of GRB~950830.
	The hardness-duration of GRB~950830 and the rest of the BATSE GRB population with published T90's is shown in Extended Data Fig. 5.
	The hardness $H_{32}$ is defined as  background counts in channel 3 (110--320 keV, green) to the counts in channel 2 (60--110 keV, yellow).
	We estimate the background by taking the mean of the bins outside the trigger region of the burst light curve.
	We find that the hardness of Pulse~A (2.09$\pm$0.10) and Pulse~B (1.83$\pm$0.11) of GRB~950830 agree within statistical errors.
	We expect two lensed pulses to exhibit the same hardness since lensing is achromatic for point sources.
	The slightly different duration between the two pulses (160ms, 130ms) is due to the lower amplitude of the second pulse (Pulse~B).
	We apply a two-component Gaussian mixture model to segregate long and short $\gamma$-ray bursts.
	The duration and hardness of GRB~950830 are typical of a short gamma-ray burst.
	We have include the autocorrelation of the light curve of GRB~950830 in Extended Data Fig. 6.

	To improve our analysis, we could \emph{a priori} constrain the magnification ratio and time delay to be the same parameter in each of the spectral channels.
	The eight parameters in the lens model -- two from each channel -- are reduced to just two shared between the four energy channels.
	However, analysing the channels separately provides an independent check that the magnification ratios and time delays are consistent, in accordance with the equivalence principle (cf. Fig. 3, Extended Data Fig. 8).
	Below, we revisit this topic, analysing the data with multiple channels simultaneously.
	Doing this four-channel nested sampling analysis on the FRED-X model with a sine-Gaussian residual becomes prohibitively expensive.
	Since we believe that this will only increase the Bayesian evidence in favour of lensing, and since our result is already quite strong ($\ln(\text{BF}) =12.9$), we leave this analysis for future work.

	Furthermore, we could include a spectral model to relate the pulse-fits in the four channels, reducing the number of free parameters in our model.
	The canonical GRB spectra model -- the Band function -- is a time averaged spectra~\cite{band_1993}, which would not suit our purposes of a time-evolving spectra.
	Addition of a spectral model requires the analysis of four time-series simultaneously, which is computationally challenging.
	We leave this as a goal for future work.

	Another future goal is to use hierarchical inference to infer the prior distributions for FRED-X parameters using the full catalogue of GRBs, the vast majority of which do not contain a lens.
	This would yield priors consistent with the population properties of GRBs.
	Again, we expect this to only strengthen the evidence in favour of lensing.
	Finally, we do not fully utilise the available high-resolution (\texttt{TTE-list} data), since this requires analysis of an 800,000 unit time series.
	This analysis is prohibitively expensive for the many-parameter models.
	We are able to run a simple FRED model Eq.~\eqref{eq:FRED}, and found that lensing was similarly favoured in all channels when running the analysis with the pre-binned \texttt{BFITS} data.

	We have thus-far assumed that we expect each image of a gravitationally lensed GRB will be statistically consistent.
	We have not discussed the many potential causes for anisotropy between the images.
	Gamma-ray bursts are highly beamed due to the ultra-relativistic velocities ($\gamma\sim10^{3-4}$) of their progenitor outflow.
	This results in the viewing angle onto the emission surface becoming important, since the radiation is beamed within and angle of $\theta\sim\gamma^{-1}$.
	The difference in viewing angle onto the source scales with the mass of the lens -- the more massive the lens, the stronger the deflection, the greater the original angular separation of the lines-of-sight onto the source.
	Assuming a homogeneous emission surface, both lines-of-sight onto the source should be viewing the same region for masses $M_\text{lens}\lesssim10^{12}$ M$_\odot$.
	For larger masses, the deflection angle becomes great enough such that the observer is viewing two emission regions which may not be in casual contact, thus the gravitationally lensed images need not be identical.
	For smaller lens masses, anisotropy in the GRB emission surface can result in the images having inherent differences.
	Finally, as discussed earlier, the  detector orientation and energy response can have a significant effect on the inferred energy spectrum, potentially resulting in a false negative identification of a gravitationally lensed pair of $\gamma$-ray bursts.

	\paragraph{Method limitations.}
	Our method provides advantages over model-agnostic approaches such as correlation, which do not include all available information, for example, Poisson counting statistics.
	Bayesian inference provides a natural framework to make quantitative statistical statements about preferred models.
	Our methodology provides a metric for detection significance, and successfully rejects dubious candidates, which trigger an autocorrelation detection (see \textbf{A rejected candidate}).
	Of course, the results of Bayesian inference are only as good as the choice of model and priors.
	We try a variety of pulse models and prior ranges to ensure our results are robust, and find that the statistically identical pulse model is consistently preferred.
	This does not preclude the existence of a better pulse model.
	We have shown that the gravitational lens candidate GRB~950830 is robustly detected using both traditional GRB lensing techniques and our Bayesian inference method.

	Finally, we point out that any method relying on self similarity can produce false positive lensing candidates if identical repeating pulses are a feature of some $\gamma$-ray bursts.
	However, we regard the “intrinsic self-similarity” hypothesis as unlikely since the vast majority of GRBs are not seen to repeat, and we cannot think of a physical mechanism that would cause a subpopulation of short GRBs to emit identical pulses.

	\paragraph{Estimate of Optical Depth.}
	A rough estimate for the optical depth to strong gravitational lensing is
	\begin{align}
		\tau = \frac{N_\text{lens}}{N_\text{GRB}} ,
	\end{align}
	where $N_\text{lens}$ is the number of multiply-imaged GRBs and $N_\text{GRB}$ is the total number of GRBs in our dataset.
	We find $N_\text{lens}=1$ lensed GRB in a dataset of $N_\text{GRB}=2,679$, so the lens probability is $P(\tau)\sim\tau= 3.7^{+7.8}_{-2.6}\times10^{-4}$ (90\% credibility), where we have used a Jeffreys prior.
	We may relate the energy density of lenses to the optical depth, $\Omega_l\sim\tau(\left<z_s\right>)$~\citelatex{Press_Gunn_1973_APJ}, where $\left<z_s\right>$ is the mean redshift of sources in the sample.

	For a point mass and a point lens, the angular Einstein radius of the lens is
	\begin{equation}
		\theta_E^2 = \frac{4GM_l}{c^2} \frac{d_A(z_l,z_s)} {d_A(z_l)d_A(z_s)}.
	\end{equation}
	Here, $G$ is the gravitational constant, $c$ the speed of light, and $M_l$ the mass of the gravitational lens.
	The angular diameter distances are defined as $d_A(z_l,z_s)=\left(\chi(z_s)-\chi(z_l)\right)/(1+z_s)$, with proper comoving distance
	\begin{equation}
		\chi(z_l,z_s) = \frac{c}{H_0}\int_{z_l}^{z_s} \frac{dz}{\sqrt{\Omega_\Lambda+\Omega_m(1+z)^3}}.
	\end{equation}
	We take $\Omega_\Lambda = 0.714$, and $\Omega_m = 0.286$, with present day Hubble constant $H_0=\unit[69.6]{km\, s^{-1} Mpc^{-1}}$.
	The angular impact parameter $\beta$ of the true position of the source to the lens can be parameterised in units of the Einstein radius, $y\equiv \beta/\theta_E$.
	Such a configuration creates two images, with time delay given by
	\begin{align}
		\Delta t (M_l, z_l, y) =
		(1+z_l) \frac{4GM_l}{c^3} f(y),
		\label{eq:timedelay2}
	\end{align}
	where
	\begin{equation}
		f(y) = \left( \frac{y}{2}\sqrt{y^2+4} + \ln{\frac{\sqrt{y^2+4}+y}{\sqrt{y^2+4}-y}}. \right)
	\end{equation}
	A source with angular impact parameter $\beta$ has an effective lensing cross-section of
	\begin{align}
		\int \sigma &= \int_{\beta_\text{min}}^{\beta_\text{max}} 2 \pi \beta \, d \beta \nonumber \\
		&= \pi \theta_E^2 \int_{y_\text{min}}^{y_\text{max}} 2y dy \nonumber \\
		&= \frac{4\pi G M_l}{c^2} \frac{d_A(z_l,z_s)} {d_A(z_l)d_A(z_s)} \int_{y_\text{min}}^{y_\text{max}} 2y \, dy .
	\end{align}
	Thus, $y_\text{min}$ and $y_\text{max}$  turn the cross-section into an annulus.
	The minimum impact parameter is set by the time delay between the arrival times of the two images.
	If the time delay is too short, the images will appear as single gamma-ray burst.
	For a point lens, we may calculate the minimum time delay for a lens of mass $M_l$ at redshift $z_l$ by inverting equation \eqref{eq:timedelay2},
	\begin{equation}
		y_\text{min}(\Delta t_\text{min}, M_l, z_l) = f^{-1}\left(\frac{c^3\Delta t_\text{min}}{(1+z_l)4GM_l}\right),
	\end{equation}
	since $f(y)$ is monotonic increasing in $y$.
	We take $\Delta t_\text{min} = \unit[10]{ms}$.
	The latter-arriving image is dimmer than the first image but must still be above the detectable flux for the detector, $\mu_2\varphi_\text{peak}>\varphi_0$, where $\mu_2$ is the magnification of the dimmer image.
	This restricts the maximum possible impact parameter~\citelatex{Blaes1992}:
	\begin{equation}
		y_\text{max}(\varphi_\text{peak}) = \left(1 + \frac{\varphi_\text{peak}}{\varphi_0} \right)^{1/4} - \left(1 + \frac{\varphi_\text{peak}}{\varphi_0}
		\right)^{-1/4},
		\label{eq:ymax}
	\end{equation}
	where $\varphi_\text{peak}/\varphi_0$ is the peak counts divided by the trigger threshold at that time.
	We estimate $y_\text{max}$ with medians of the $\varphi_\text{peak}/\varphi_0$ for the peak flux on $\unit[64]{ms}$, $\unit[256]{ms}$, and $\unit[1024]{ms}$ integrations from the BATSE C$_\text{max}$ table: \url{https://gammaray.nsstc.nasa.gov/batse/grb/catalog/4b/tables/4br_grossc.cmaxmin}, which are 1.5, 2.2, and 2.5 respectively.
	The final cross section is then
	\begin{equation}
		\sigma(\Vec{x}) = \frac{4\pi G M_l}{c^2} \frac{d_A(z_l)d_A(z_l,z_s)} {d_A(z_s)}
		\left(y_\text{max}^2-y_\text{min}^2\right)
		\Theta\left(y_\text{max}-y_\text{min}\right)
	\end{equation}
	where $\Vec{x} \equiv (M_l, z_l, z_s, \varphi_\text{peak}, \varphi_0, \Delta t_\text{min})$ and $\Theta$ is the Heaviside step function.

	The optical depth is the number density $n(z_l)$ of lenses at redshift $z_l$, multiplied by the effective lensing cross-section of each lens $\sigma(\vec{x})$, integrated over $z\in(0,z_s)$:
	\begin{equation}
		\tau(\vec{x}) = \frac{1}{d\Omega} \int_0^{z_s} dV(z_l) n(z_l) \int d\sigma (\vec{x}) .
	\end{equation}
	We assume a constant comoving density of lenses,
	\begin{equation}
		n_l(z_l)=n_0(1+z_l)^3.
	\end{equation}
	The number density of lenses can be related to their energy density $\Omega_l$ through
	\begin{equation}
		n_0 = \frac{\rho_c\Omega_l}{M_l} = \frac{3H_0^2\Omega_l}{8\pi G M_l}
		\label{eq:numberdensity}.
	\end{equation}
	With comoving volume element
	\begin{align}
		dV(z) &= \chi^2(z) \frac{d \chi(z)}{dz}  \,dz d\Omega \nonumber \\
		&= \chi^2(z) \frac{c}{H_0} \frac{dz \, d\Omega}{\sqrt{\Omega_\Lambda + \Omega_m (1+z)^3}},
	\end{align}
	we have
	\begin{align}
		\tau(\vec{x}) &=\frac{3H_0\Omega_l}{2 c \chi(z_s)} \int_0^{z_s} dz_l \frac{(1+z_l)\chi(z_l)}{\sqrt{\Omega_\Lambda + \Omega_m (1+z_l)^3}}[\chi(z_s)-\chi(z_l)] \nonumber \\
		&\quad \cdot \left[
		y^2_\text{max}(\varphi_\text{peak},\varphi_0) - y^2_\text{min}(\Delta t_\text{min}, M_l, z_l)\right] .
		\label{eq:opticaldepth}
	\end{align}
	With an estimated lens probability of $P(\tau)\sim\tau\approx3.7\times10^{-4}$, we infer the lens density by inversion of Eq.~\eqref{eq:opticaldepth}.
	The result of this integral is shown in Extended Data Fig. 9 for several mean source redshifts $\left<z_s\right>$ each with the three permutations of $\varphi_\text{peak}/\varphi_0$ from the BATSE C$_\text{max}$ table.

	The 7 BATSE bursts with known redshifts are GRB~970508: $z=0.835$, GRB~970828: $z=0.958$, GRB~971214: $z=3.412$, GRB~980425: $z=0.0085$, GRB~980703: $z=0.967$, GRB~990123: $z=1.600$, GRB~990510: $z=1.619$, with mean $\left<z\right>=1.34$.
	The average spectroscopic redshift of Swift $\gamma$-ray bursts is $\left<z\right>=2.2$~\cite{Xiao2011ApJ}.
	We are unable to accurately estimate the redshift of GRB~950830 or the BATSE catalogue in general due to the inherent degeneracy between the effects of cosmological redshift and relativistic beaming on a $\gamma$-ray burst light curve.
	We argue that a mean BATSE GRB redshift of $\left<z_s\right>\sim2$ is appropriate based on the redshifts of known BATSE bursts and the spectroscopically determined redshifts of other GRB catalogues.
	We include the derived energy densities for a number of redshifts in Extended Data Fig. 10 for the comparison.
	The lens densities are calculated from the source redshifts and optical depth through Eq.~\eqref{eq:opticaldepth}.
	With the inferred lens densities $\Omega_l$, we may exclude our calculated globular cluster density $\Omega_\text{gc}\sim8\times10^{-6}$ based on Poisson statistics.
	Observing one gravitational lens in $\sim2,679$ light curves is very unlikely for such a low cosmological density.

	The present day number density is given by
	\begin{align}
		n_\textsc{imbh}(z_l,z_s) = \frac{\Omega_\textsc{imbh}(z_s) \, \rho_c}
		{M_\textsc{imbh}(z_l)} ,
	\end{align}
	where $\rho_c$ is the critical energy density of the Universe and $M_\textsc{imbh}(z_l)$ is the mass of the lens.
	With $\Omega_\textsc{imbh}(\left<z_s\right>\sim2) = 4.6^{+9.8}_{-3.3}\times10^{-4}$, and $(1+z_l)M_\textsc{imbh} = 5.5^{+1.7}_{-0.9}\times 10^4 $ M$_\odot\implies M_\textsc{imbh}(z_l\sim1)\sim 2.8^{+1.7}_{-0.9}\times 10^4 $ M$_\odot$ yields $n_\textsc{imbh}=\unit[2.3^{+4.9}_{-1.6}\times10^3]{Mpc^{-3}}$ through Eq.~\eqref{eq:numberdensity}.
	Where $z_l\sim1$ comes from a gravitational lens being most likely to occur halfway between the observer and source.
	Uncertainty on the density of IMBHs arises from the fact that $N_\text{lens}$ is Poisson distributed.
	We calculate the uncertainty on $n_l$ assuming $z_l=0$, as we do not know where the lens is, only that its most probable redshift is $z_l\sim1$.
	We ignore the uncertainty due to the lens mass as it is much more precisely determined.
	In order to calculate the uncertainty on $n_\textsc{imbh}$, we assume that the likelihood of the data given $n_\textsc{imbh}$ follows an $N=1$ Poisson distribution.
	Employing a Jeffreys' prior,
	\begin{align}
		\pi(n_\textsc{imbh}\propto n_\textsc{imbh}^{-1/2} ),
	\end{align}
	we obtain a 90\% credible interval of $n_\textsc{imbh}=\unit[0.7-7.2\times10^3]{Mpc^{-3}}$.
	The calculated lens densities for each redshift are summarised in Table~\ref{tab:opticaldepths}.

	\begin{table*}[htb]
		\begin{center}
			\begin{tabular}{l l l l}
				$\left<z_s\right>$ & $\Omega_l(z_s)$ & $n_l(z_l,z_s)$ & $\Omega_\text{gc}$ exclusion \\\hline
				0.1 & $3.5^{+7.4}_{-2.5}\times10^{-1}$ & $7.9^{+17.0}_{-5.6}\times10^{5}$ & 99.99999\% \\
				0.5 & $1.2^{+2.5}_{-0.8}\times10^{-2}$ & $3.6^{+7.7}_{-2.6}\times10^{4}$ & 99.9988\% \\
				1.0 & $2.4^{+5.2}_{-1.7}\times10^{-3}$ & $9.1^{+19.0}_{-6.5}\times10^{3}$ & 99.988\% \\
				1.34 & $1.2^{+2.6}_{-0.9}\times10^{-3}$ & $5.1^{+11.0}_{-3.6}\times10^{3}$ & 99.97\% \\
				2.0 & $4.6^{+9.8}_{-3.3}\times10^{-4}$ & $2.3^{+4.9}_{-1.6}\times10^{3}$ & 99.85\% \\
				5.0 & $4.2^{+8.9}_{-3.0}\times10^{-5}$ & $3.6^{+7.7}_{-2.6}\times10^{2}$ & 95.05\% \\
			\end{tabular}
		\end{center}
		\caption{
			{\bf The inferred lens densities $\Omega_l$ for mean source redshift $z_s$.}  A median peak counts ratio $\widetilde{C}= 2.5$ from the BATSE C$_\text{max}/$C$_\text{min}$ table for 1,024ms integration times is assumed.
			The peak count ratios are defined through $C\equiv C_\text{max}/C_\text{min}$.
			$C_\text{max}$ is the maximum detected counts over a given integration period.
			$C_\text{min}$ is the minimum number of counts that would trigger the second most brightly illuminated detector at that time. Further details are given in the Supplementary Section calculation of optical depths.
		}
		\label{tab:opticaldepths}
	\end{table*}

	\paragraph{Magnification bias.}
	It is possible that magnification bias may affect the estimate of the expected probability of lensing events.  The possibility of magnification bias is discussed in \citelatex{Blaes1992}, but the details are  updated here from \cite{Paciesas1999}.
	In order for magnification bias to be greater than a few percent, the cumulative number counts $dN$ of the physical parameter $P$ by which the events are detected needs to have $\alpha > 2$ where $dN \propto P^{-\alpha}$.
	In the case of GRBs, $P$ is the peak flux over the trigger energy range.  For BATSE, the number counts as a function of peak flux are given in Fig. ~6~\cite{Paciesas1999}. Estimating the value of $\alpha$ from these plots gives $\alpha \sim 1.1$.  The trigger flux for GRB~950830 falls near the faint end of the distribution at $2.81\pm 0.12$ $\gamma$ cm$^{-2}$ sec$^{-1}$ in the $50-300$ keV energy range over integration times of $\unit[64]{ms}$, $\unit[256]{ms}$ and $\unit[1024]{ms}$.

	\paragraph{The lens is unlikely to be a globular cluster.}
	Let us assume that all globular clusters have the same mass as the discovered deflector.
	The globular cluster mass function has a turnover at $\sim2\times10^5$ M$_\odot$~\cite{jordan_2007}, so this approximation is not necessarily a bad one, especially given the uncertainty in the inferred lens mass.
	The Milky Way formed from an overdensity on a scale of approximately 20 Mpc$^3$.
	If the 200 globular clusters in the Milky Way is an average number count for a given cosmic volume, then the number density of globular clusters is approximately $n\sim 10 $~Mpc$^{-3}$. Thus with $n\sim10$ Mpc$^{-3}$, $M\sim10^5$ M$_\odot$:
	\begin{align*}
		\Omega_\textsc{gc} &= \rho \frac{8\pi G}{3 H_0^2}, \qquad \rho = nM  \\
		&\approx 8\times10^{-6}
	\end{align*}
	This is too low to be consistent with our inferred value of $\Omega\sim4\times10^{-4}$, suggesting that the lens is not a globular cluster.
	Assuming a mean source redshift of $\left<z_s\right>\sim2$ in the BATSE catalogue, we exclude a globular cluster lens at $99.85\%$ credibility.
	See Extended Data Fig. 10 for exclusion credibilities for different mean redshifts.
	The mean of BATSE gamma-ray bursts with known redshift $\left<z\right>=1.34$ gives an exclusion credibility of $99.97\%$, with $\Omega_l=1.43\times10^{-3}$.

\begin{figure}
		\centering
		\includegraphics[width=1\textwidth]{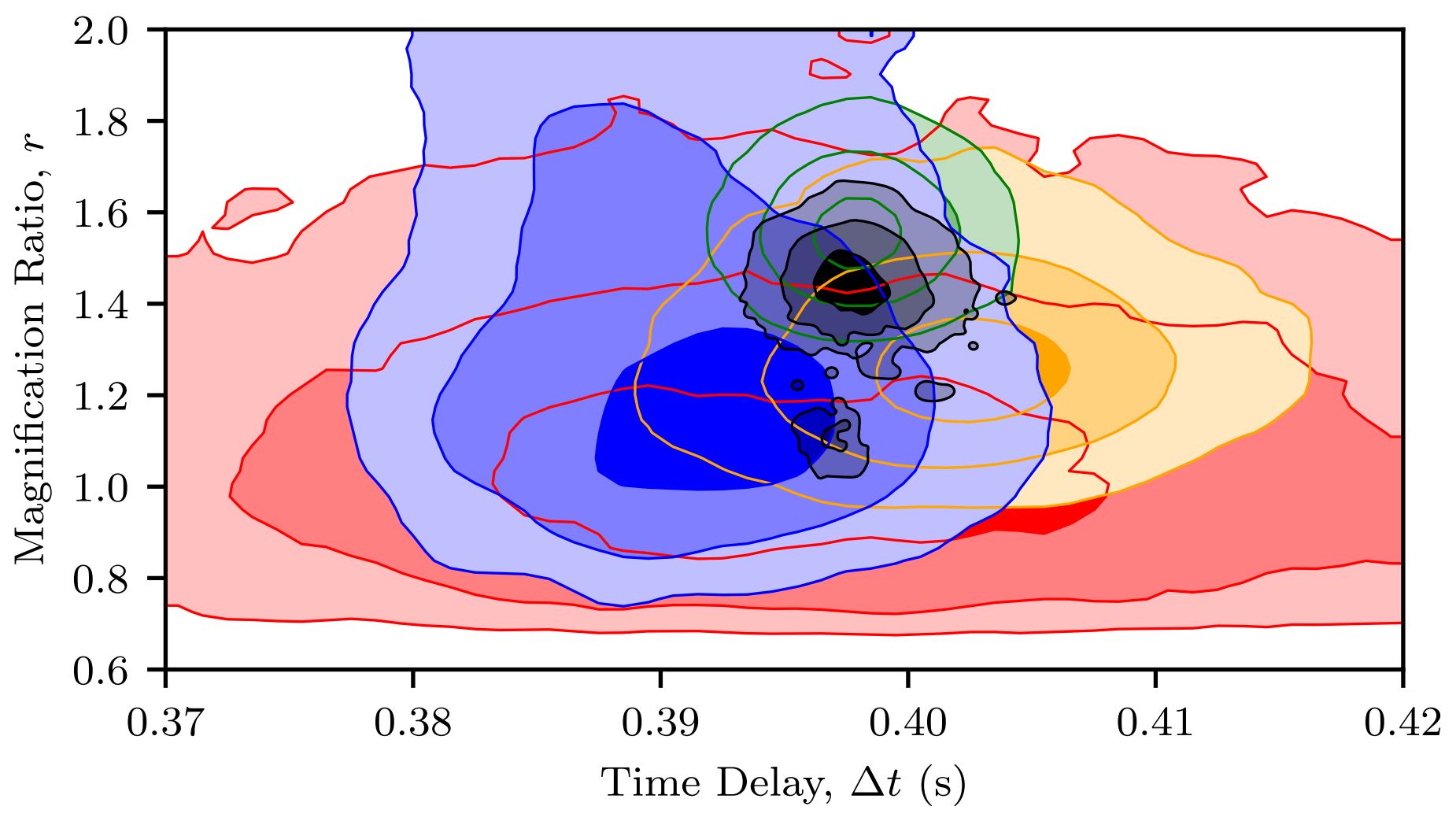}
		\caption{{\bf
				Marginalised posterior distribution of time-delays vs magnification ratios for the gravitationally lensed $\gamma$-ray burst GRB~950830.}
			The colours indicate the following energy channels, red: $20-60$ keV, yellow: $60-110$ keV, green: $110-320$ keV, blue: $320-2,000$ keV.
			The black region is the result of the product of Gaussian kernel density estimates for each of the four channels, resulting in tighter constraints on the time delay and magnification ratio.
			The contours contain 39.3\%. 86.4\%, and 98.9\% of the probability density.
		}
		\label{fig:delmudelt}
	\end{figure}

	\paragraph{Combining data from different channels.}
	In the analysis presented above we analyse each channel independently.
	This enables us to carry out posterior checks (see Fig.~\ref{fig:delmudelt}) while controlling computational costs.
	However, in order to estimate the significance of the lensing signal, we combine the results from each channel together, increasing the resolving power to obtain a single Bayes factor for the lensing versus no-lensing hypotheses.
	There are several steps.
	First, we take the posterior samples used to calculate the credible intervals for delay time $\Delta t$ and magnification ratio $r$ in Fig.~\ref{fig:delmudelt} and use a kernel density estimator (KDE) to obtain an analytic description of the posterior distribution for each channel $i$:
	\begin{align}
		p(\Delta t, r | d_i) .
	\end{align}
	This step is necessary in order to obtain smooth functions that can be multiplied together.
	Next, we invoke Bayes' theorem to convert the KDEs to likelihoods distributions for each channel:
	\begin{align}
		{\cal L}(d_i | \Delta t, r) = \frac{{\cal Z}_i}{\pi(\Delta t, r)}
		p(\Delta t, r | d_i) .
	\end{align}
	If the two pulses are created by a lens, then the delay time $\Delta t$ and magnification ratio $r$ are the same for each channel.
	Thus, the total evidence for the lensing hypothesis is given by the product of likelihoods for each data channel, marginalised over $(\Delta t, r)$:
	\begin{align}\label{eq:ZS_tot}
		\mathcal{Z}_\text{lens, tot} = &
		\int d\Delta t \int dr \, \pi(\Delta t, r) \prod_i {\cal L}(d_i | \Delta t, r) \\
		%%%%%%%%%%%%%%
		= & \int d\Delta t \int dr \, \pi(\Delta t, r) \prod_i \left(\frac{\mathcal{Z}_\text{lens, i}}{\pi(\Delta t, r)} p_i(\Delta t, r|d_i) \right)  .
		\label{eq:evidenceintegralproduct}
	\end{align}
	On the other hand, if the two pulses are not created from a lens, there is no reason for the delay time and magnification ratio to be the same for each channel, and so the total evidence for the null hypothesis is given simply by the product of the evidence values for each channel:
	\begin{align}\label{eq:ZN_tot}
		{\cal Z}_\text{null}^\text{tot} = \prod_i {\cal Z}_\text{null, i}
	\end{align}
	Evaluating Eqs.~\ref{eq:ZS_tot} and~\ref{eq:ZN_tot}, we obtain a total Bayes factor $\ln(\text{BF})=12.93$ in favour of the lensing hypothesis.

	\paragraph{False alarm probability.}
	In Bayesian statistics, we carry out model selection using the posterior odds~\citelatex{intro},
	\begin{equation}
		\mathcal{O}^\text{lens}_\text{null} = \frac{\mathcal{Z}_\text{lens}}{\mathcal{Z}_\text{null}} \frac{\pi_\text{lens}}{\pi_\text{null}},
	\end{equation}
	where $\mathcal{Z}_\text{lens} / \mathcal{Z}_\text{null}=\text{BF}$ is the Bayes factor from Eq.~\eqref{eq:bayes}.
	The prior odds, $\pi_\text{lens}/\pi_\text{null}$, expresses the relative probability assigned \emph{a priori} to the hypothesis of lensing over a null model.
	If we assign an equal prior probability to both hypotheses, then
	\begin{align}
		\mathcal{O}^\text{lens}_\text{null} &= \frac{\mathcal{Z}_\text{lens}}{\mathcal{Z}_\text{null}} \\
		&= \frac{p_\text{lens} }{1 - p_\text{lens}} ,
	\end{align}
	and so the false alarm probability is
	\begin{align*}
		1 - p_\text{lens} &= \frac{1}{1+\text{BF}} \\
		&= \frac{1}{1+e^{12.9}} \\
		&= 2.5\times 10^{-6}
	\end{align*}

	A more conservative approach is to assign prior odds equal to the reciprocal of the total number of bursts searched (or equivalently, equal to the optical depth), which gives us:
	\begin{align*}
		\mathcal{O}^\text{lens}_\text{null} &= \frac{\mathcal{Z}_\text{lens}}{\mathcal{Z}_\text{null}}
		\frac{\pi_\text{lens}}{\pi_\text{null}} \\
		&= \frac{\text{BF}}{2,679}, \\
	\end{align*}
	which gives a false alarm probability of
	\begin{align*}
		1 - p_\text{lens} &= \frac{1}{1+\text{BF}/2,679} \\
		&= \frac{1}{1 + e^{12.9}/2,679} \\
		&= 6.5 \times 10^{-3}.
	\end{align*}
	We see that the strong preference for the lensing hypothesis persists even taking into account trial factors.

	\paragraph{A rejected candidate.}
	For illustrative purposes, we also include one gravitational lensing candidate identified by the autocorrelation algorithm, which our Bayesian framework strongly rejects.
	The light curve for GRB~911031 is shown in Extended Data Fig. 7 as the sum of the four BATSE large area detector broadband energy channels and separately for each channel in the first and third panels respectively.
	Each colour indicates a different energy channel, red: $\unit[20-60]{keV}$, yellow: $\unit[60-110]{keV}$, green: $\unit[110-320]{keV}$, blue: $\unit[320-2,000]{keV}$.
	The light curve has the sort of repeating pulse structure which one might na\"ively mistake for lensing.
	The time delay is even similar in each energy channel.
	The second and fourth panels show the correlogram (autocorrelation) of the summed and spectral light curves respectively.

	The results of autocorrelation indicate very strongly that there is some similar structure in the two pulses.
	However, the pulse modelling prefers a two-pulse model in every channel because the two pulses are not precise duplicates.
	Not only are the pulse shapes ($\tau, \xi$) different between the two pulses, but the time-delays and magnification ratios ($\Delta t, r$) inferred through parameter estimation are inconsistent.
	Extended Data Fig. 8  shows the gravitational lens parameter posterior distributions for lens model fit to GRB~911031.
	Gravitational lensing for point sources is achromatic, so the time-delay and magnification ratio posteriors for each channel should be overlapping for a gravitational lensing candidate.
	The clear correlation between energy, magnification, and time delay immediately suggests that these are two independent pulses of a GRB and not a gravitational lensing echo.
	The $\ln(\text{BF})$ in favour of a two-pulse model are 229.0, 301.7, 374.0 15.4 for channels 1, 2, 3, and 4 respectively.
	We can therefore firmly rule out the lensing interpretation for this burst.

	\section*{Data Availability}
	The BATSE data catalogue is available from the NASA data archive: \url{https://heasarc.gsfc.nasa.gov/FTP/compton/data/batse/trigger}.
	We use the \texttt{discsc}, \texttt{tte}, and \texttt{tte\_list} datatypes in our search.
	The data used in our analysis of GRB~950830 is found at \url{https://heasarc.gsfc.nasa.gov/FTP/compton/data/batse/trigger/03601_03800/03770_burst/tte_bfits_3770.fits.gz}, \url{https://heasarc.gsfc.nasa.gov/FTP/compton/data/batse/trigger/03601_03800/03770_burst/tte_list_3770.fits.gz}.

	\section*{Code Availability}
	The analysis code \texttt{PyGRB}~\cite{Paynter2020} has been written in \texttt{python}~\cite{python3} by J.P. and is freely available at \url{https://github.com/JamesPaynter/PyGRB} under the BSD 3-Clause License.
	\texttt{PyGRB} is built around Monash University's \texttt{Bilby} nested sampling wrapper, with additional \texttt{FITS} I/O functionality provided by \texttt{AstroPy}~\cite{astropy2013}.
	The software uses the \texttt{NumPy}~\cite{numpy} and \texttt{SciPy}~\cite{2020SciPy-NMeth} computational libraries.
	Plotting makes use of the \texttt{Matplotlib}~\cite{Hunter2007} and \texttt{corner}~\cite{corner} libraries.

	\FloatBarrier

	\begin{figure}
		\centering
		\includegraphics[width=1\textwidth]{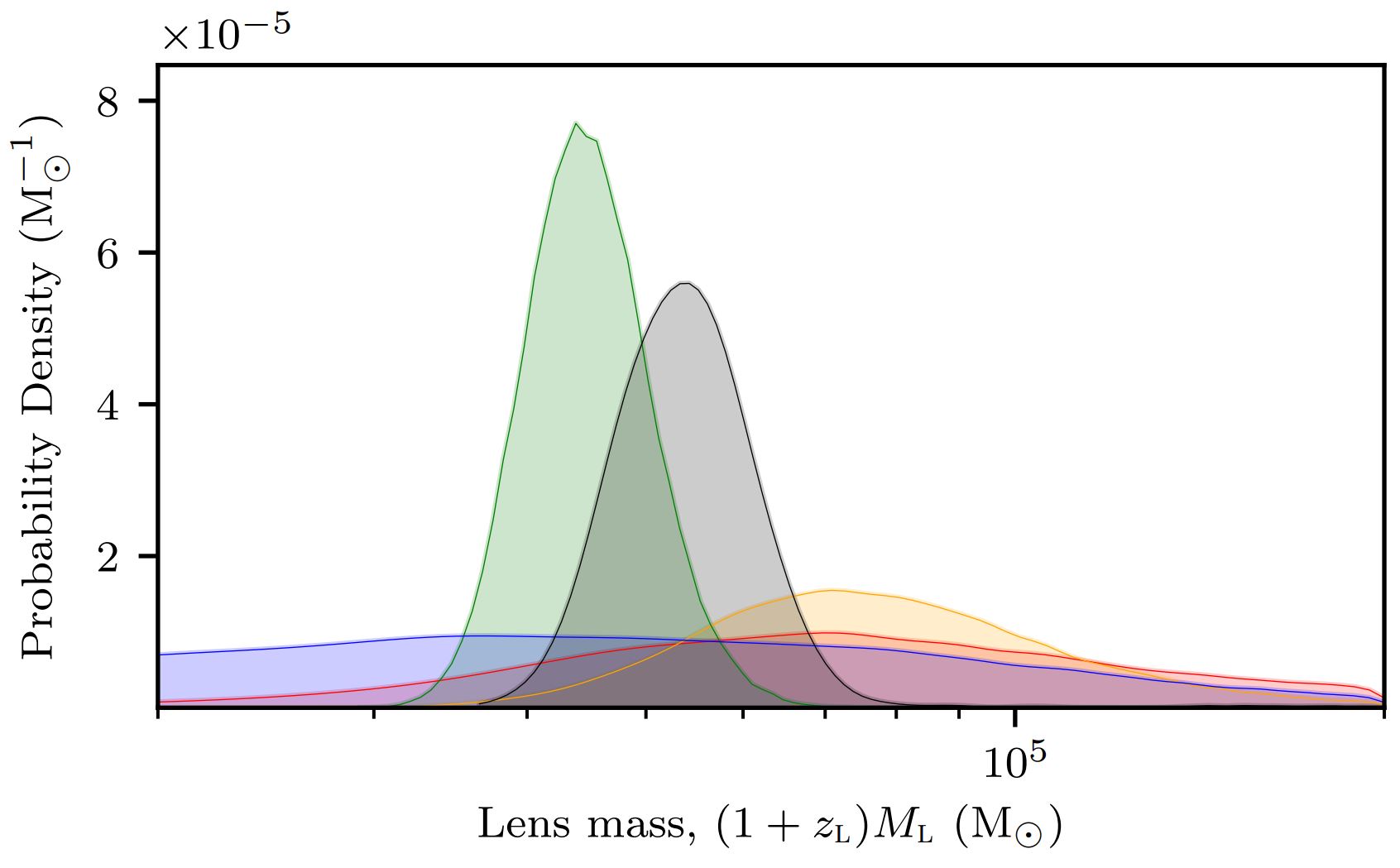}
		\caption{ {\bf The redshifted lens mass.}
			Taking the time delays and magnification ratios posterior samples shown in Figure  (\ref{fig:delmudelt}), the mass of the gravitational lens may be determined through equation \eqref{eq:mass_redshift}.
			The colours indicate the mass recovered from the following energy channels, red: $20-60$ keV, yellow: $60-110$ keV, green: $110-320$ keV, blue: $320-2,000$ keV.
			The black histogram is the mass inferred from the Gaussian kernel density estimate of the four energy channels (see Figure (\ref{fig:delmudelt})).
			The median of the mass is $(1 + z_\text{l}) M_\text{l} = 5.5^{+1.7}_{-0.9}\times 10^4 $ M$_\odot$ (90\% credibility).
			The inferred mass is dependent on the redshift of the lens itself.
			Without any knowledge of the lens or source redshift, all we are able to say is $0<z_\text{l} \leq z_\text{s}$, and so $M_\text{l}\leq (1 + z_\text{l}) M_\text{l}$.
		}
		\label{fig:mass_dist}
	\end{figure}

	\begin{figure}
	    \centering
	    \includegraphics[width=.8\textwidth]{"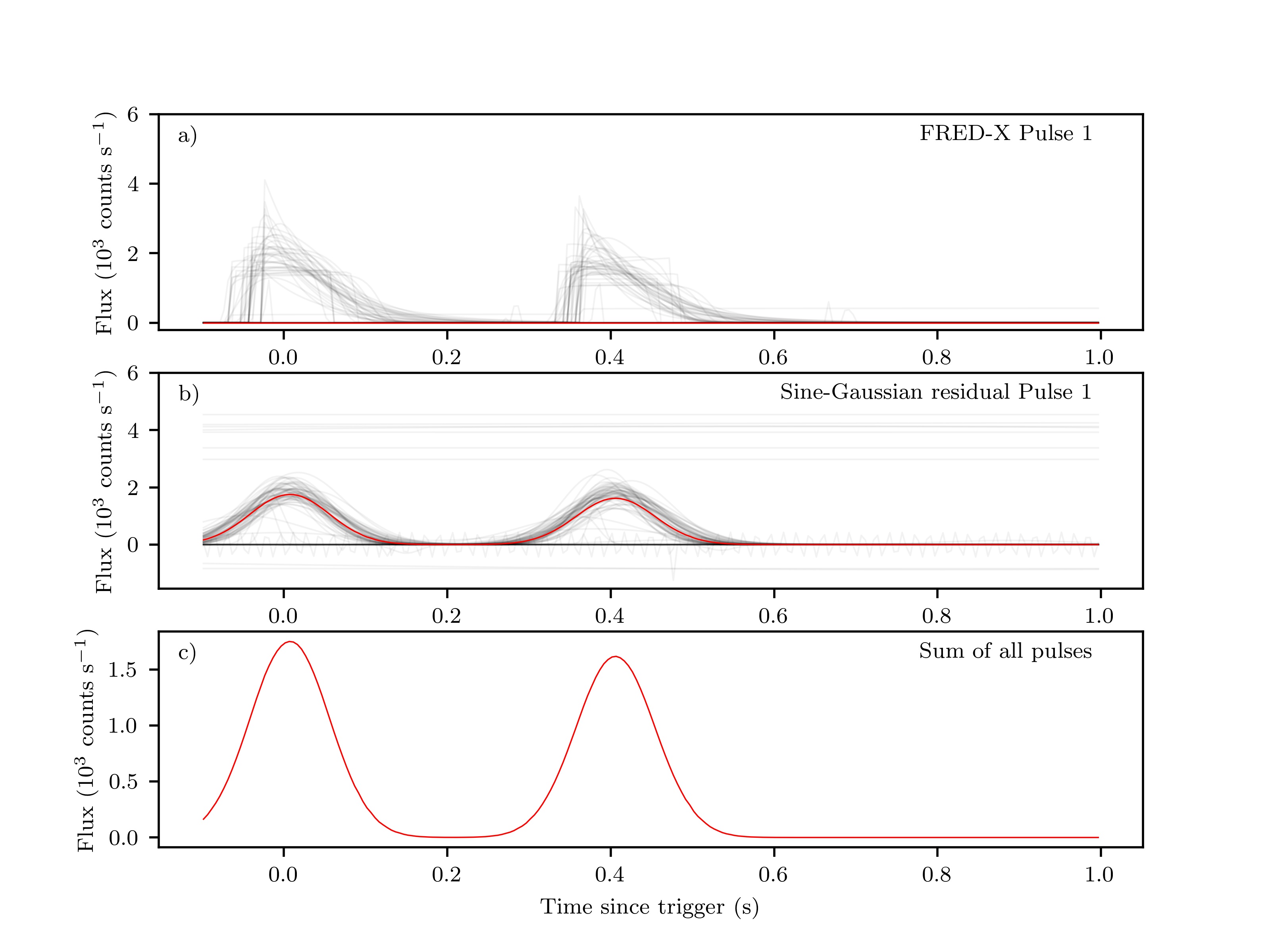"}
	    \caption{{\bf
	    The individual pulses that make up channel 1 (red: $20-60$ keV) of Figure \ref{fig:lightcurve}.}
	    a) The solid red curves are the median of $\sim60,000$ FRED-X pulses sampled from the posterior distributions.
	    Two hundred of these curves are sampled and shown in black.
	    b) The same as a) for the sine-Gaussian residual.
	    c) The sum of the medians of the pulses in a) and b).
	    }
	    \label{fig:lines1}
	\end{figure}

	\begin{figure}
	    \centering
	    \includegraphics[width=.8\textwidth]{"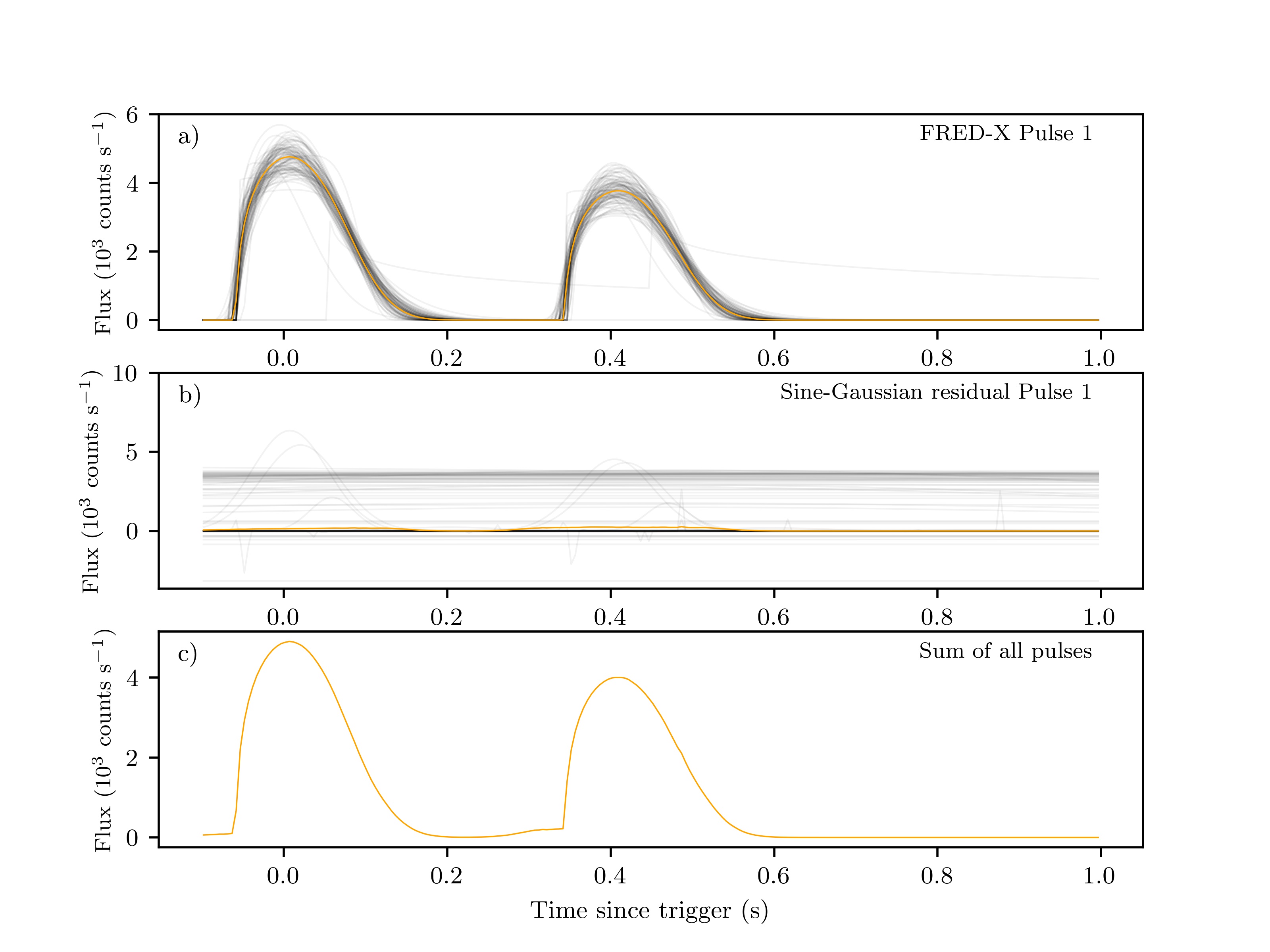"}
	    \caption{{\bf
	    The individual pulses that make up channel 2 (yellow: $60-110$ keV) of Figure \ref{fig:lightcurve}.}
	    a) The solid yellow curves are the median of $\sim60,000$ FRED-X pulses sampled from the posterior distributions.
	    200 of these curves are sampled and shown in black.
	    b) The same as a) for the sine-Gaussian residual.
	    c) The sum of the medians of the pulses in a) and b).
	    }
	    \label{fig:lines2}
	\end{figure}

	\begin{figure}
	    \centering
	    \includegraphics[width=.8\textwidth]{"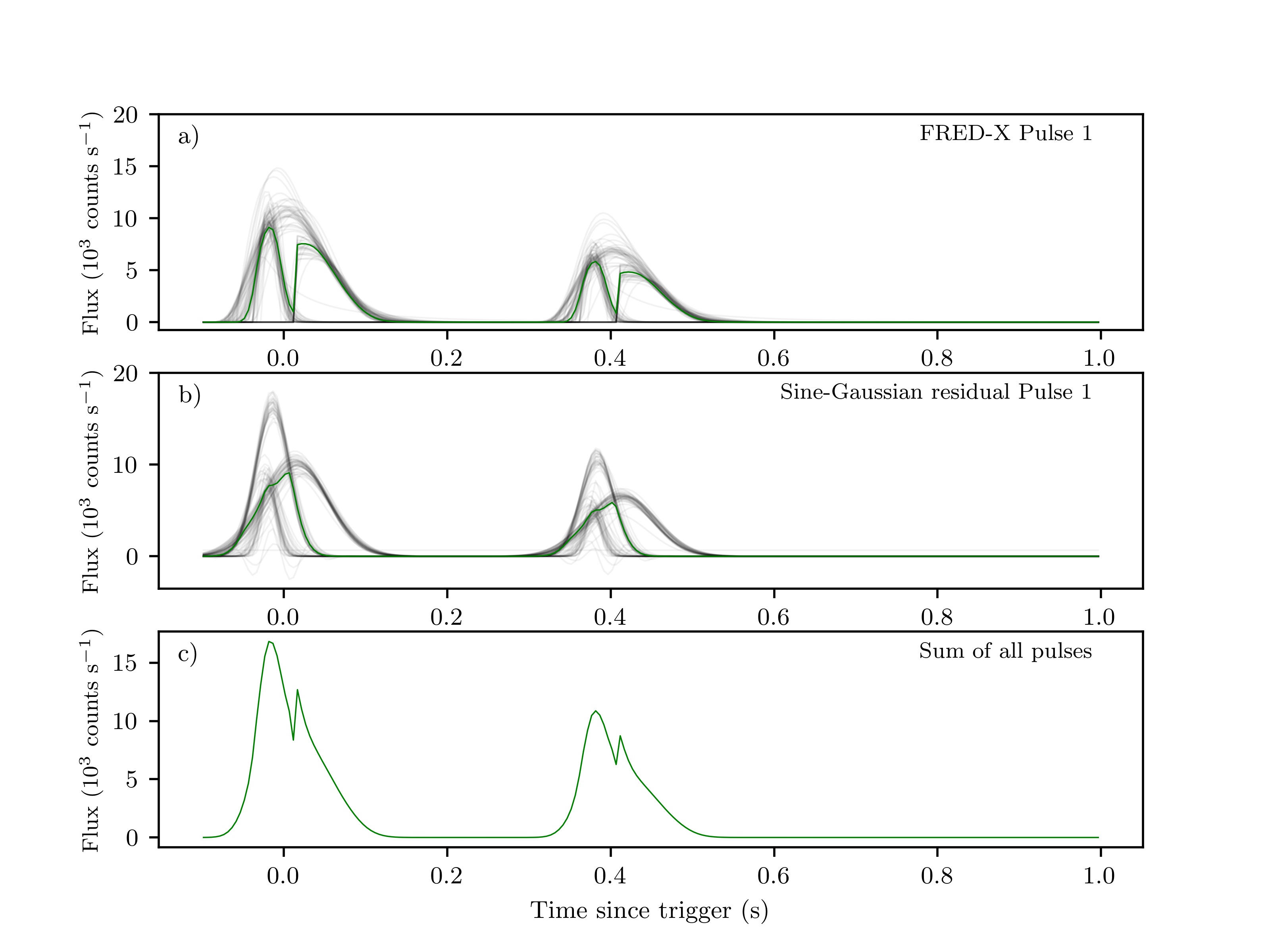"}
	    \caption{{\bf
	    The individual pulses that make up channel 3 (green: $110-320$ keV) of Figure \ref{fig:lightcurve}.}
	    a) The solid green curves are the median of $\sim60,000$ FRED-X pulses sampled from the posterior distributions.
	    200 of these curves are sampled and shown in black.
	    b) The same as a) for the sine-Gaussian residual.
	    c) The sum of the medians of the pulses in a) and b).
	    }
	    \label{fig:lines3}
	\end{figure}

	\begin{figure}
	    \centering
	    \includegraphics[width=.8\textwidth]{"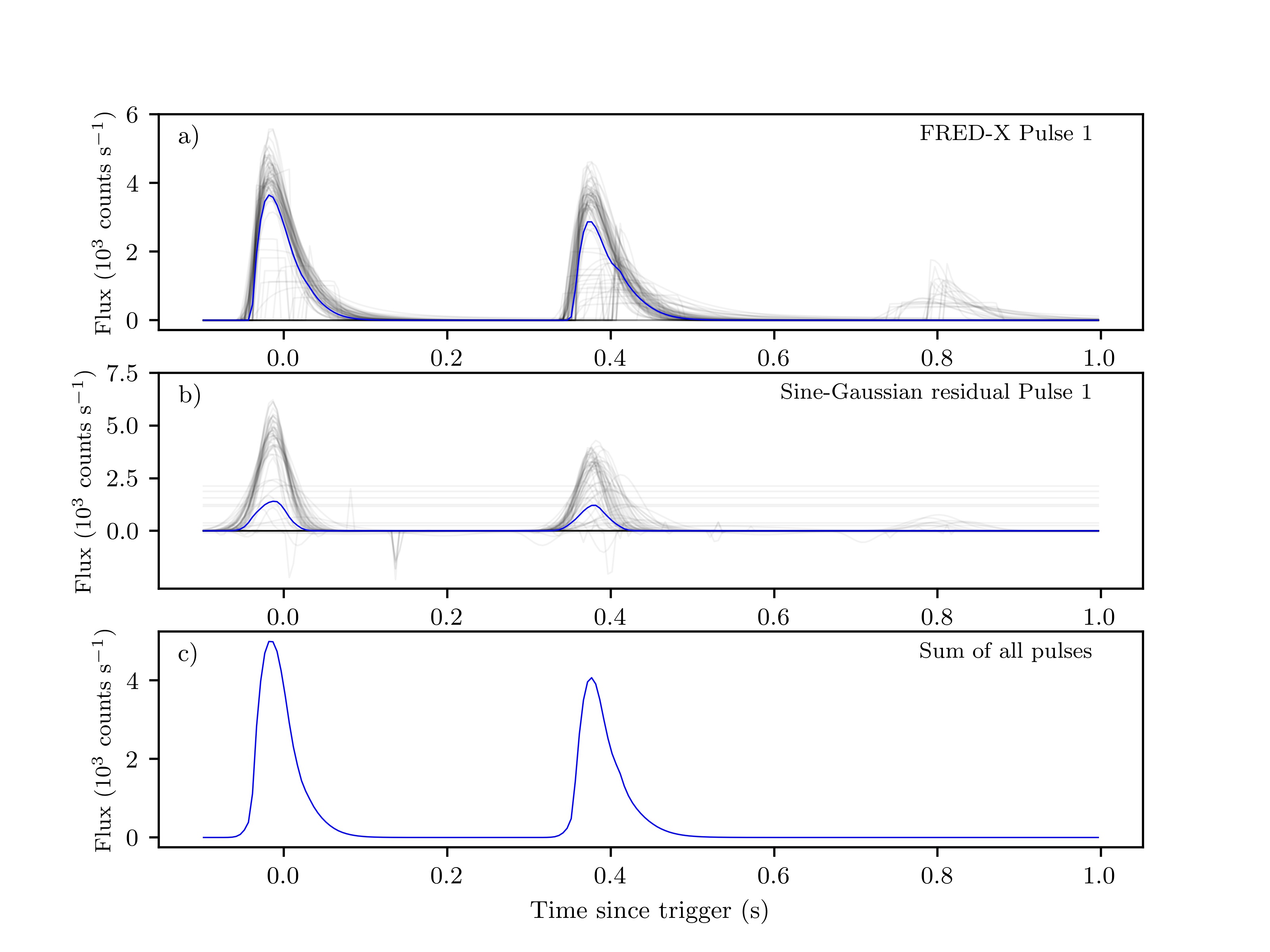"}
	    \caption{
	    {\bf
	    The individual pulses that make up channel 4 (blue: $320-2,000$ keV) of Figure \ref{fig:lightcurve}.
	    }
	    a) The solid blue curves are the median of $\sim60,000$ FRED-X pulses sampled from the posterior distributions.
	    200 of these curves are sampled and shown in black.
	    b) The same as a) for the sine-Gaussian residual.
	    c) The sum of the medians of the pulses in a) and b).
	    }
	    \label{fig:lines4}
	\end{figure}

	\begin{figure}[htb!]
	    \centering
	    \includegraphics[width=\textwidth]{"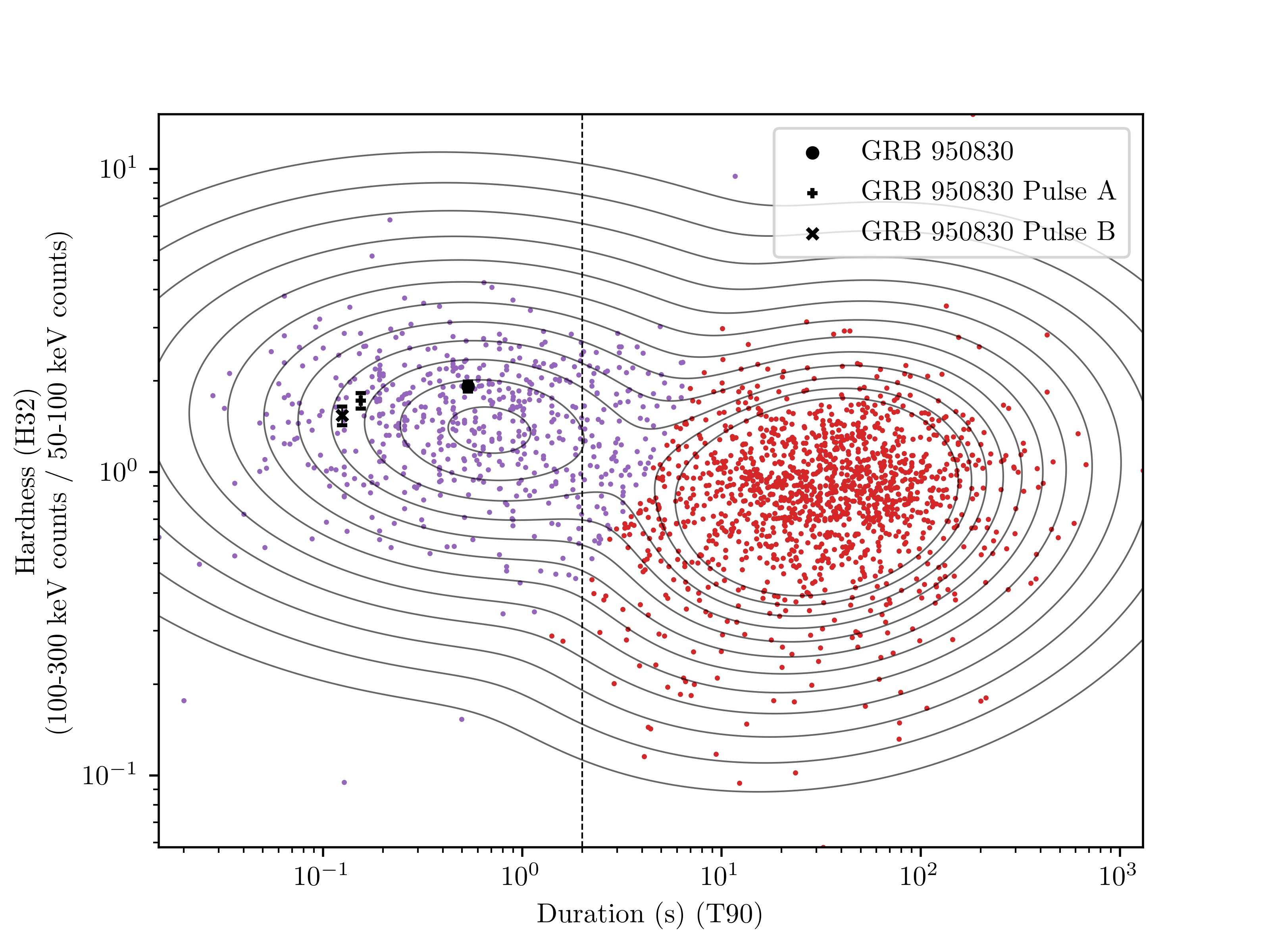"}
	    \caption{
	    {\bf The Hardness-Duration plot of BATSE GRBs.}
	    The T90 durations are taken from the BATSE data tables: \url{https://gammaray.nsstc.nasa.gov/batse/grb/catalog/4b/index.html}.
	    We calculate the hardness ratios for each of the GRBs with a listed T90.
	    The short $\gamma$-ray burst population is shown in purple, and the long GRB population in red.
	    Iso-likelihood contours of a two-component Gaussian mixture model are plotted in grey.
	    The plotted uncertainties in the hardness ratio are defined by 1-$\sigma$ statistical errors on the number of counts in the numerator and denominator.
	    }
	    \label{fig:hardness}
	\end{figure}

	\begin{figure}
	    \centering
	    \includegraphics[width=1\textwidth]{"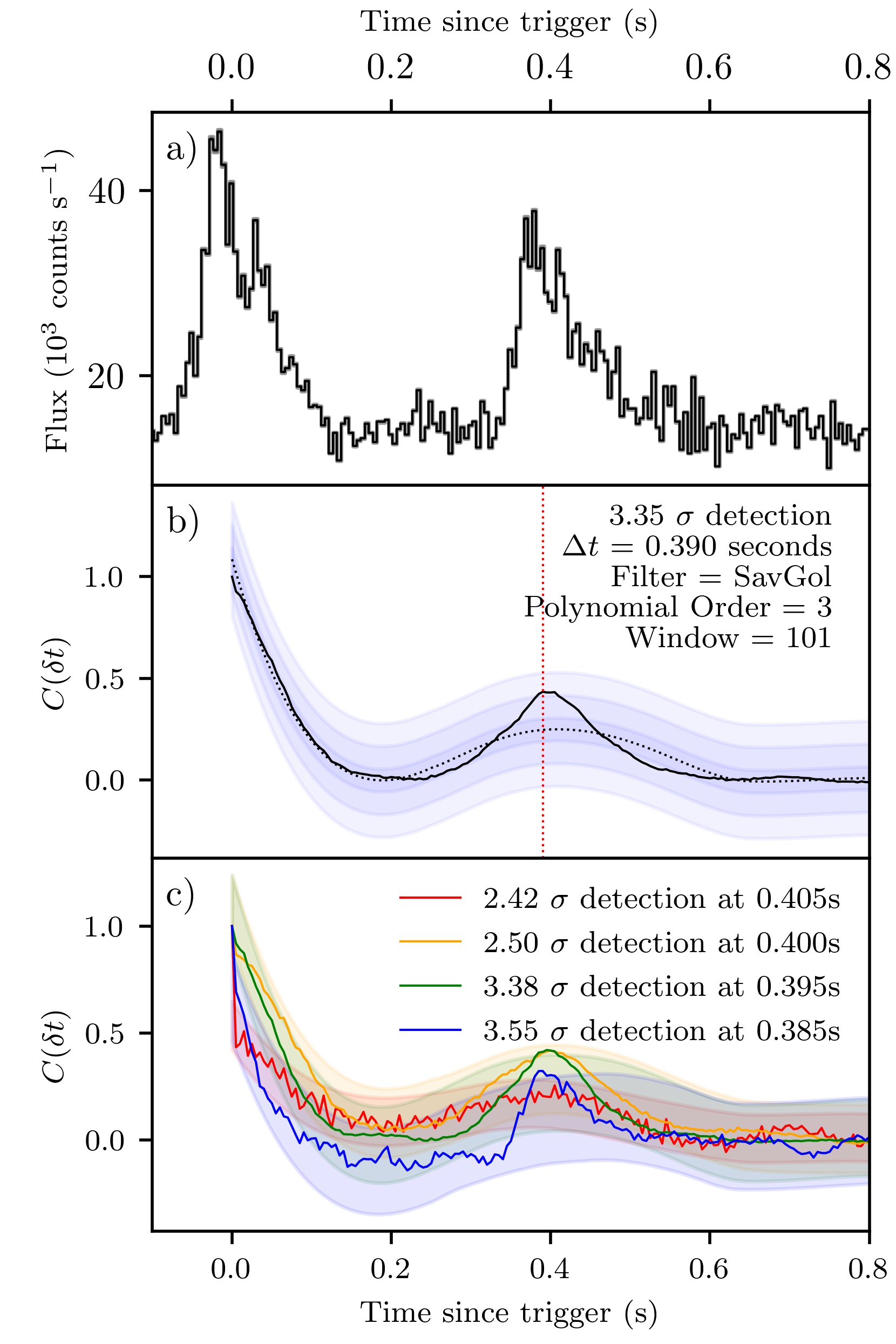"}
	    \caption{{\bf
	    The autocorrelation of the light curve of GRB~950830.}
	    a) The sum of the four energy channels, $\sim20-2,000$ keV.
	    b) The autocorrelation function of the summed light curve, where the autocorrelation is defined in equation \eqref{eq:correlation}.
	    The black dotted line is a fit to the light curve with a 3rd order Savitzky-Golay smoothing filter with a 101 bin smoothing window.
	    The vertical red dotted line is the point of maximum deviation between the ACF and the Savitzky-Golay smoothing filter at $\delta t = 0.390$ seconds.
	    The blue shaded regions delineate regions of 1-$\sigma$, 3-$\sigma$, and 5-$\sigma$ away from the Savitzky-Golay fit.
	    The dispersion between the autocorrelation function and the fit, $\sigma^2$, is defined in equation \eqref{eq:dispersion}.
	    c) The autocorrelation function for each of the 4 BATSE large area detector broadband energy channels.
	    Each colour indicates a different energy channel, red: $20-60$ keV, yellow: $60-110$ keV, green: $110-320$ keV, blue: $320-2,000$ keV.
	    The shaded regions delineate $3-\sigma$ deviance from the Savitzky-Golay fits, which are omitted for clarity.
	    }
	    \label{fig:autocorrelation3770}
	\end{figure}

	\begin{figure}[htb!]
	    \centering
	    \includegraphics[width=\textwidth]{"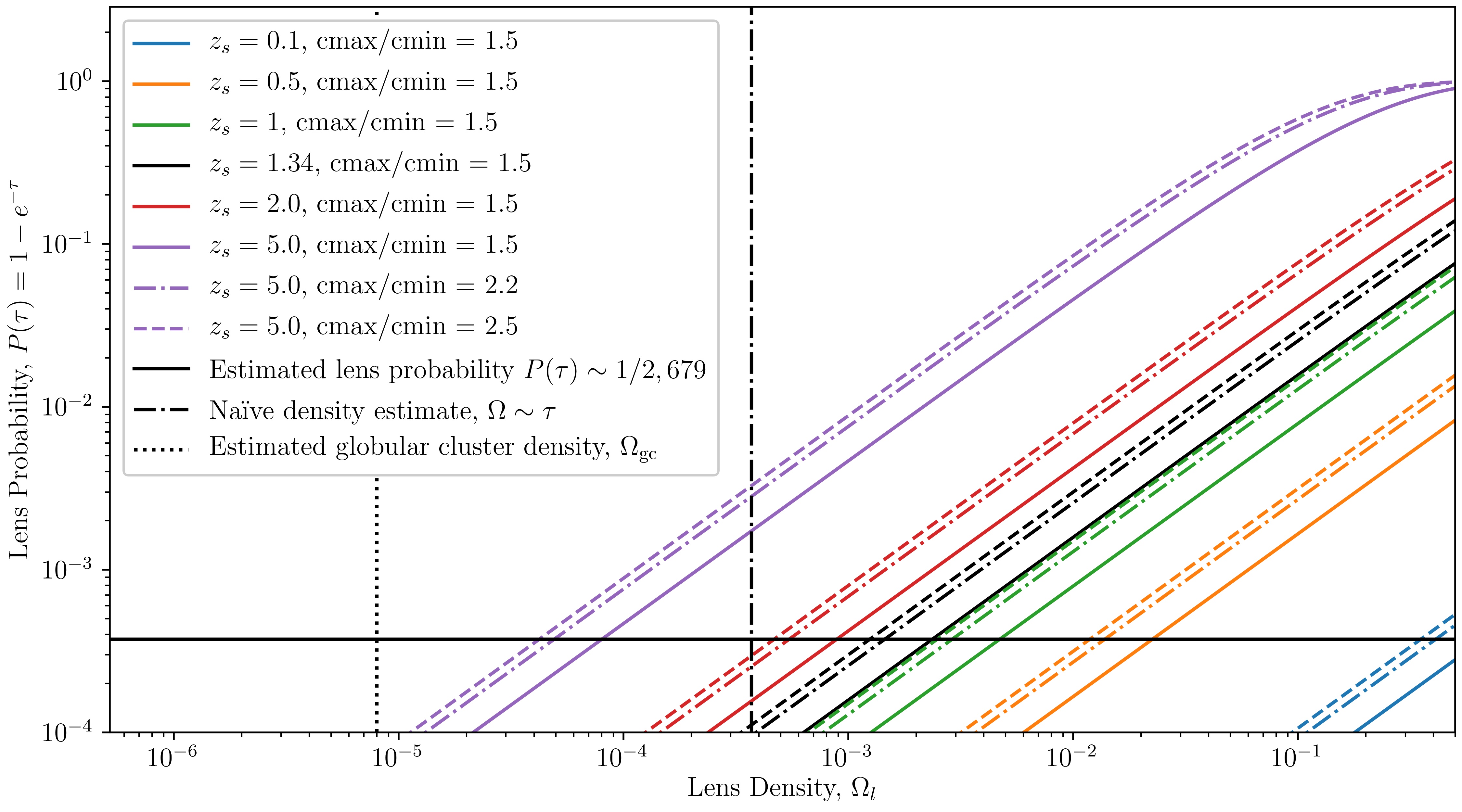"}
	    \caption{{\bf Optical depth as a function of source redshift $z_s$.}
	    We estimate the optical depth for mean source redshifts $z_s=0.1$: blue, $z_s=0.5$: orange, $z_s=1.0$: green, $z_s=1.34$: black, $z_s=2.0$: red, $z_s=5.0$: purple based on Eq.~\eqref{eq:opticaldepth}.
	    The median $C_\text{max}/C_\text{min}$ values of 1.5, 2.2, and 2.5 taken as the magnification limit cutoff (Eq.~\eqref{eq:ymax}) are shown as solid, dash-dot, and dashed curves respectively.
	    The solid black horizontal line is the estimate lens probability based on seeing one event in $\sim2,700$ light curves.
	    The dotted black vertical line is the estimated globular cluster density, $\Omega_\textsc{gc}$.
	    The dash dot vertical black line is the na\"ive estimate for the density $\Omega_\text{lens}\sim\tau$.
	    The calculated lens densities for each redshift are summarised in Table~\ref{tab:opticaldepths}.
	    }
	    \label{fig:opticaldepths}
	\end{figure}

	\begin{figure}
	    \centering
	    \includegraphics[height=.8\textheight]{"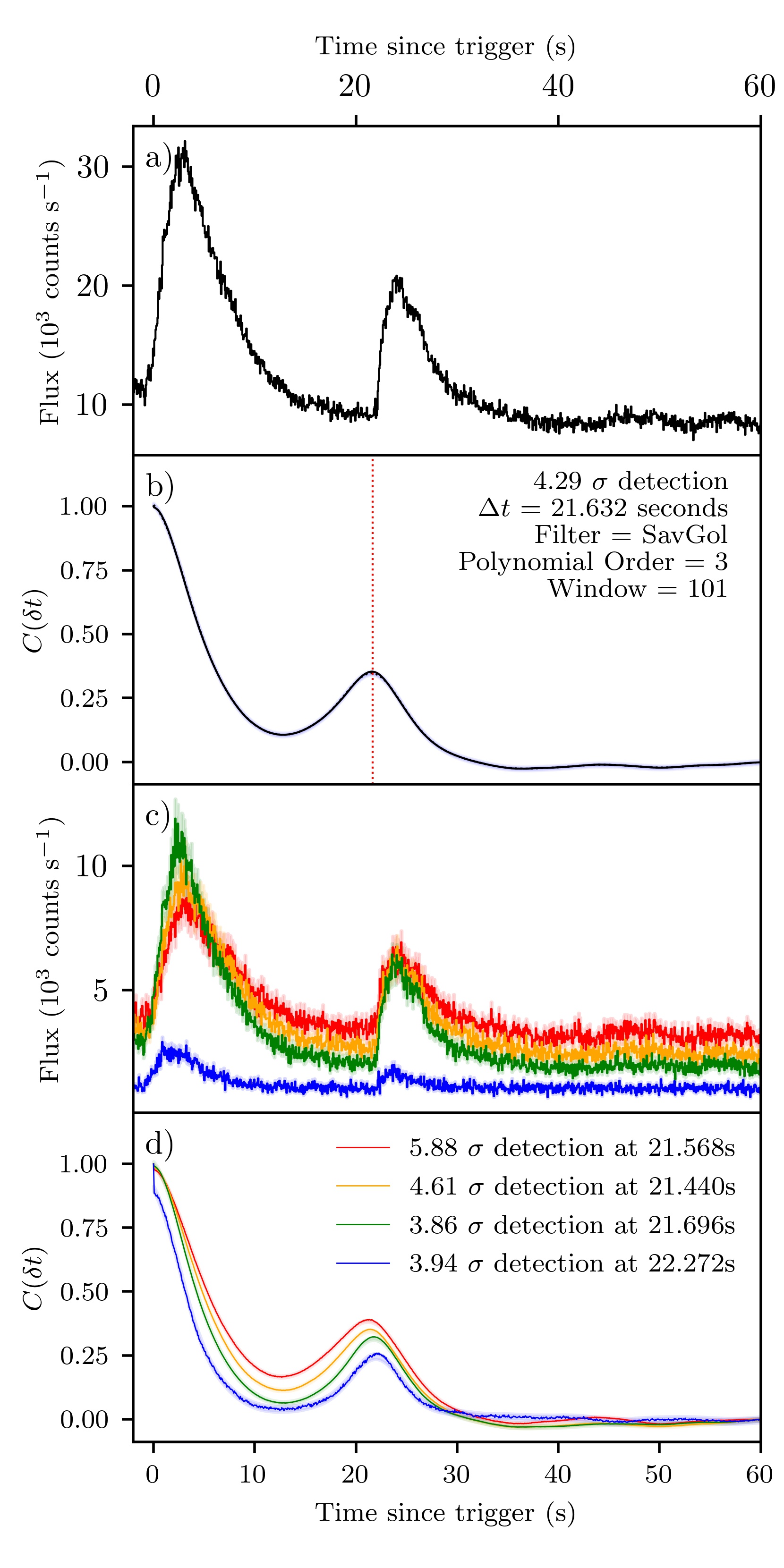"}
	    \caption{{\bf
	    The autocorrelation of the light curve of GRB~911031.}
	    a) The sum of the four energy channels, $\sim20-2,000$ MeV.
	    b) The autocorrelation function of the summed light curve, where the autocorrelation is defined in equation \eqref{eq:correlation}.
	    The dotted line is a fit to the light curve with a 3rd order Savitzky-Golay smoothing filter with a 101 bin smoothing window.
	    The blue shaded regions delineate regions of 1-$\sigma$, 3-$\sigma$, and 5-$\sigma$ away from the Savitzky-Golay fit.
	    The dispersion between the autocorrelation function and the fit, $\sigma^2$, is defined in equation \eqref{eq:dispersion}.
	    c) The light curve for each of the 4 BATSE large area detector broadband energy channels.
	    Each colour indicates a different energy channel, red: $20-60$ keV, yellow: $60-110$ keV, green: $110-320$ keV, blue: $320-2,000$ keV.
	    d) The autocorrelation function for each light curve channel.
	    The shaded regions delineate $3-\sigma$ deviance from the Savitzky-Golay fits, which are omitted for clarity.
	    }
	    \label{fig:autocorrelation973}
	\end{figure}

	\begin{figure}
	    \centering
	    \includegraphics[width=.8\textwidth]{"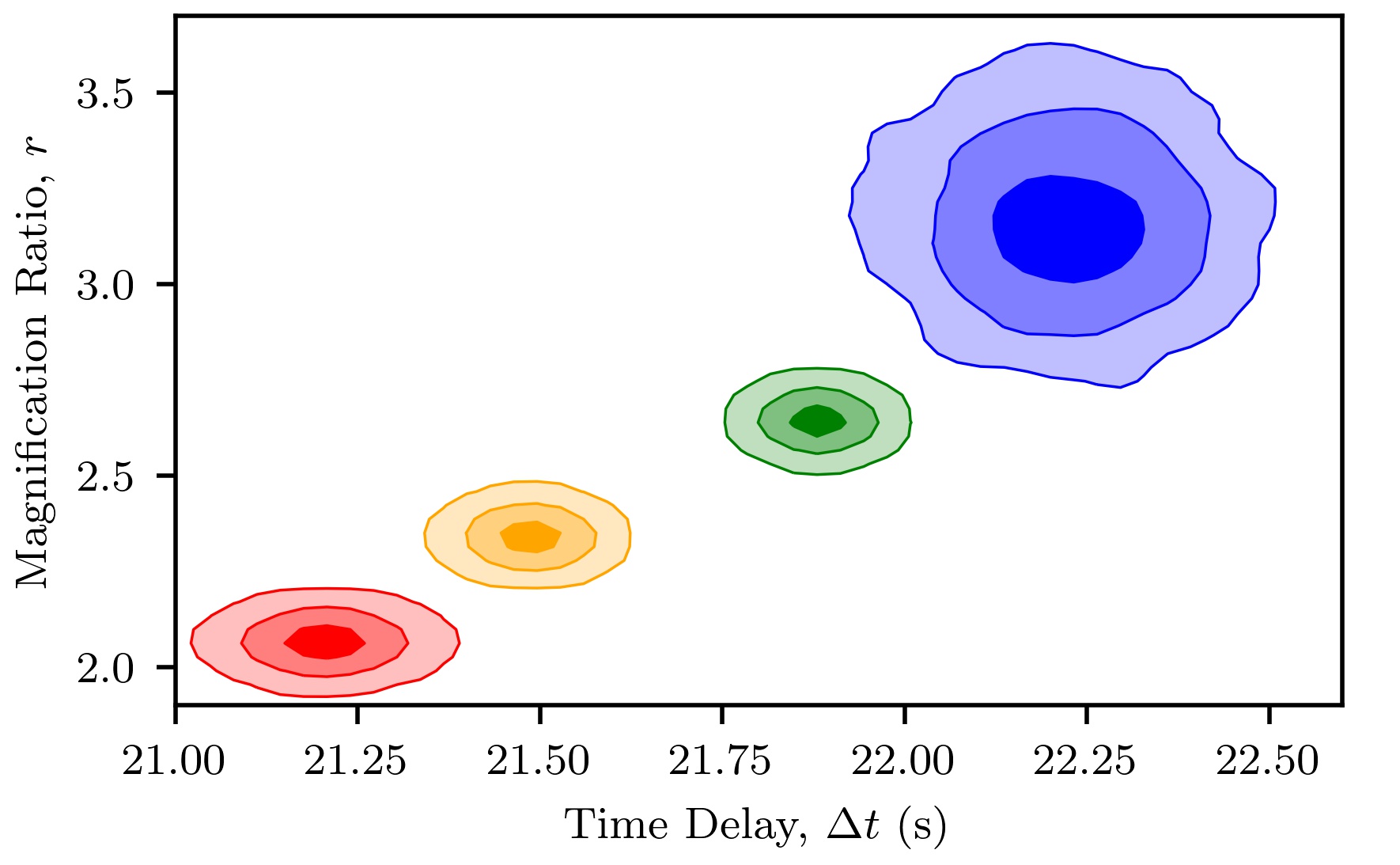"}
	    \caption{{\bf
	    The gravitational lens parameter posterior distributions for a model fit to GRB 911031 for each of the 4 BATSE large area detector broadband energy channels.}
	    Each colour indicates a different energy channel, red: $20-60$ keV, yellow: $60-110$ keV, green: $110-320$ keV, blue: $320-2,000$ keV.
	    Contours contain 39.3\%. 86.4\%, and 98.9\% of the probability density.
	    The light curve of GRB 911031 is shown in Figure \ref{fig:autocorrelation973}.
	    }
	    \label{fig:delmudelt973}
	\end{figure}

\end{document}